\documentclass[preprint,pteplogo]{ptephy}%%%%%% to generate preprint number with ptep logo

\preprintnumber{KUNS-2476 / YITP-14-2} %%% Insert preprint number here

\usepackage{color}
\usepackage{amsmath}
\usepackage{graphicx}
\usepackage{amssymb}
\usepackage{bm}

\newcommand{\Comments}[1]{}
\newcommand{\be}{\begin{equation}}
\newcommand{\ee}{\end{equation}}
\newcommand{\ba}{\begin{align}}
\newcommand{\ea}{\end{align}}
\newcommand{\nn}{\nonumber}

\newcommand{\VEV}[1]{\langle{#1}\rangle}
\newcommand{\chibar}{{\bar{\chi}}}

\newcommand{\Fint}{{\cal D}}

\newcommand{\Tr}{\mathrm{Tr}}
\newcommand{\bk}{\mathbf{k}}
\usepackage[normalem]{ulem}  % \sout{old text} for strikeout
\newcommand{\com}[1]{{\color[rgb]{0,0,1}{#1}}}
\newcommand{\comm}[1]{{\color[rgb]{1,0,1}{#1}}}

\renewcommand\sout{\bgroup \color{red} \ULdepth=-.5ex \ULset}
\newcommand\soutb{\bgroup \color{blue} \ULdepth=-.5ex \ULset}
\newcommand{\PSfig}[2]{\includegraphics[width=#1]{#2}}
\newcommand{\vk}{{\bold{k}}}

\newcommand{\vkt}{{\vk,\tau}}
\newcommand{\vkmt}{{-\vk,\tau}}
\newcommand{\vkbt}{{\bar{\vk},\tau}}
\newcommand{\vkmbt}{{-\bar{\vk},\tau}}
\newcommand{\expv}[1]{\left< #1 \right>}
\newcommand{\mrm}[1]{\mathrm{#1}}
\newcommand{\Nt}{N_{\tau}}
\newcommand{\equref}[1]{Eq.~(\ref{#1})}
\newcommand{\secref}[1]{Sec.~\ref{#1}}
\newcommand{\figref}[1]{Fig.~\ref{#1}}

\newcommand\soutg{\bgroup \color{green} \ULdepth=-.5ex \ULset}

\renewcommand{\sout}[1]{}
\renewcommand{\comm}[1]{{\color[rgb]{0,0,0}{#1}}}
\renewcommand{\com}[1]{{\color[rgb]{0,0,0}{#1}}}

\begin{document}

\title{Auxiliary field Monte-Carlo simulation\\ 
of strong coupling lattice QCD 
for QCD phase diagram }

\author{
\name{Terukazu Ichihara}{1,2,\ast}, 
\name{Akira Ohnishi}{2,\dag}, 
and \name{Takashi Z.~Nakano}{3,\ddag}%\thanks{These authors contributed equally to this work}
}
%%%%%%%%%%% The \name command should be used as \name{Insert author name here}{Insert affiliation number here}
%%%%% Please use \thanks for contributed author details

%%%%%%%%%%% The \affil command should be used as \affil{Insert affiliation number here}{Insert author address here}
\address{\affil{1}{Department of Physics, Kyoto University,
         Kyoto 606-8502, Japan}
\affil{2}{Yukawa Institute for Theoretical Physics, Kyoto University,
         Kyoto 606-8502, Japan}
\affil{3}{Kozo Keikaku Engineering Inc.,
         Tokyo 164-0012, Japan}
\email{t-ichi@ruby.scphys.kyoto-u.ac.jp}, 
{\dag}ohnishi@yukawa.kyoto-u.ac.jp, 
{\ddag}takashi-nakano@kke.co.jp
}

\begin{abstract}%

We study the QCD phase diagram in the strong coupling limit
with fluctuation effects
by using the auxiliary field Monte-Carlo method.
We apply the chiral angle fixing technique
in order to obtain finite chiral condensate
in the chiral limit in finite volume.
The behavior of order parameters suggests that chiral phase transition is the second order or crossover at low chemical potential and the first order at high chemical potential. 
Compared with the mean field results,
the hadronic phase is suppressed at low chemical potential, 
and is extended at high chemical potential
as already suggested
in the monomer-dimer-polymer simulations.
We find that the sign problem
originating from the bosonization procedure
is weakened by the phase cancellation mechanism;
a complex phase from one site tends to be
canceled by the nearest neighbor site phase
as long as low momentum auxiliary field contributions dominate.

\end{abstract}

\subjectindex{ D30, %Quark matter
      B64, %Lattice QCD
      D34, % Lattice QCD calculations in nuclear physics
      }%Quark deconfinement, quark-gluon plasma production, and phase transitions (see also 12.38.Mh Quark-gluon plasma in quantum chromodynamics; 21.65.Qr Quark matter in nuclear matter)

\maketitle
\section{Introduction}
\label{sec:Intro}
%QCD phase diagram
%The phase diagram of Quantum Chromodynamics (QCD) 
%is one of the most interesting topics 
%in both quark-hadron physics and astrophysics. 

Quantum Chromodynamics (QCD) phase diagram is attracting much attention in recent years.
At high temperature ($T$), 
there is a transition from quark-gluon plasma (QGP) to hadronic matter via the crossover transition, which 
was realized in the early universe and is now extensively studied
in high-energy heavy-ion collision experiments at RHIC and LHC. 
At high quark chemical potential ($\mu$), we also expect 
the transition from baryonic to quark matter, 
which may be realized in cold dense matter such as the neutron star core.
Provided that the high density transition is the first order,
the QCD critical point (CP) should exist as the end point
of the first order phase boundary.
Large fluctuations of the order parameters 
around CP may be observed 
in the beam energy scan program at RHIC.

The Monte-Carlo simulation of the lattice QCD (MC-LQCD)
is one of the first principle non-perturbative methods 
to investigate the phase transition. 
We can obtain various properties of QCD:
hadron masses and interactions,
color confinement,
%, spontaneous breaking hadron mass
chiral and deconfinement transitions,
equation of state,
and so on.
%This indicate that the critical temperature at $\mu =0$ is $T_{c} \simeq 160-190 (\mathrm{MeV})$. 
We can apply MC-LQCD to the low $\mu$ region,
but not to the high $\mu$ region because of the notorious sign problem.
The fermion determinant becomes complex at finite $\mu$,
then the statistical weight is reduced by the average phase factor
$\expv{e^{i\theta}}$, where $\theta$ is the complex phase
of the fermion determinant.
There are many attempts to avoid the sign problem
such as the reweighting method~\cite{Reweighting},
the Taylor expansion method~\cite{Taylor}, 
the analytic continuation from imaginary chemical potential~\cite{ImagMu},
the canonical ensemble method~\cite{Canonical}, 
the fugacity expansion~\cite{Fugacity},
the histogram method~\cite{Histogram},
and the complex Langevin method~\cite{Langevin}.
Many of these methods are useful for $\mu/T < 1$, 
while it is difficult to perform the Monte-Carlo simulation
in the larger $\mu$ region.

Recent studies suggest that CP may not be reachable
in phase quenched simulations~\cite{NoGo}:
In the phase quenched simulation for $N_f=2$,
the sampling weight at finite $\mu$ is given as
$|\det D(\mu)|^2=\det D(\mu)(\det D(\mu))^*=\det D(\mu) \det D(-\mu^*)$,
where $D$ represents the fermion matrix for a single flavor.
The phase quenched fermion determinant
for real quark chemical potential $\mu_d=\mu_u=\mu \in \mathbb{R}$
%the above phase quenched weight
is the same as that
for finite isospin and vanishing quark chemical potentials, $\mu_d=-\mu_u=\mu$.
Thus the phase quenched phase diagram in the temperature-quark chemical potential $(T,\mu)$ plane
would be the same as that in the temperature-isospin chemical potential $(T, \delta\mu)$ plane,
as long as we can ignore the mixing of $u$ and $d$ condensates.
%\sout{as justified in the large $N_c$ case}.
We do not see any critical behavior in the finite $\delta\mu$ simulations
outside of the pion condensed phase~\cite{Finiteisospin}.
By comparison, \com{the} pion condensed phase appears at large $\delta \mu$, where the above correspondence does not apply.
We may have CP
inside the corresponding pion condensed phase.
However, we now have an overlap problem; gauge configurations in the pion condensed phase 
would be very different from those of compressed baryonic matter which we aim to investigate.
Therefore, we need to find methods other than the phase quenched simulation
in order to directly sample appropriate configurations
in cold dense matter 
for the discussion of CP and the first order phase transition.

The strong coupling lattice QCD (SC-LQCD) is one of the methods 
to study finite $\mu$ region based on the strong coupling expansion 
($1/g^2$ expansion) of the lattice QCD.
There are some merits to investigate \com{the }QCD phase diagram 
using SC-LQCD, 
while the strong coupling limit (SCL)
is the opposite limit of the continuum limit.
First,
the effective action is given in terms of color singlet components, 
then we expect suppressed complex phases of the fermion determinant
and a milder sign problem. 
We obtain the effective action by integrating out the spatial link variables 
before the fermion field integral.
This point is different from the standard treatment of MC-LQCD, 
in which we integrate out the fermion field 
before the link integral.
Second, 
we can obtain insight into
\com{the }QCD phase diagram 
from the mean-field studies at strong coupling.
The chiral transition has systematically and analytically been studied in the strong coupling expansion ($1/g^2$ expansion) under the mean-field approximation:
the strong coupling limit 
(leading order, $\mathcal{O}(1/g^0)$)~\cite{SCL,large_d,MF-SCL,Faldt,BilicDeme,Bilic,NLOchiral,NLOPD,NNLO,Jolicoeur},
the next-to-leading order 
(NLO, $\mathcal{O}(1/g^2)$)~\cite{NLOchiral,NLOPD,Bilic,BilicDeme,Faldt,Jolicoeur,NNLO}, 
and the next-to-next-to-leading order
(NNLO, $\mathcal{O}(1/g^4)$)~\cite{Jolicoeur,NNLO} 
with staggered fermion. 

It is necessary 
to go beyond the mean-field treatment and
to include the fluctuation effects
of the order parameters for quantitative studies
of the finite density QCD.
Monomer-dimer-polymer (MDP) simulation is one of the methods beyond the mean-field approximation.
We obtain the effective action of quarks after the link integral, 
and evaluate the fermion integral by summing up
monomer-dimer-polymer configurations~\cite{KarschMutter}.
The phase diagram shape is modified to some extent,
compared with the mean-field results on an isotropic lattice:
the chiral transition temperature is reduced by 10-20 \% at $\mu=0$,
and the hadronic phase expands to higher $\mu$ direction 
by 20-30 \%~\cite{MDP,PhDFromm,MDPsign}.
Until now, we can perform MDP simulations
only in the strong coupling limit, $1/g^2=0$, 
and the finite coupling corrections are evaluated in the reweighting method~\cite{SC-Rewei}.
Since both finite coupling and fluctuation effects are important
to discuss the QCD phase diagram,
we need to develop a theoretical framework which includes
both of these effects.
%%%%%%%%%%%%%%%%%%%%%%%%%%%%%%%%%%%%%%%%%%%%%%%%%5

In this work, we study the QCD phase diagram by using 
an auxiliary field Monte-Carlo (AFMC) method 
as a tool to take account of the fluctuation effects of the auxiliary fields.
AFMC is widely used 
in nuclear many-body problems~\cite{AFMC-NP,AFMC-NPCMP}
and in condensed matter physics such as ultra cold atom 
systems~\cite{AFMC-NPCMP}.
In AFMC, we introduce the auxiliary fields to decompose the fermion 
interaction terms
and carry out the Monte-Carlo integral of 
auxiliary fields, which is assumed to be static and constant 
in the mean-field approximation.
We can thus include the fluctuation effects of the auxiliary fields 
in AFMC beyond the mean-field approximation.

Another important aspect of this paper is how to fix the chiral angle,
the angle between the \com{zero momentum }scalar and pseudoscalar modes. 
In finite volume, 
symmetry of the theory is not broken spontaneously and an order parameter, in principle, vanishes.
In spin systems, 
a root mean square order parameter is applied to obtain the appropriate order parameter~\cite{MCsim1}.
We here use a similar method, chiral angle fixing (CAF).
The lowest momentum modes of auxiliary fields, 
$\sigma_0$ and $\pi_0$ in Eqs. \eqref{Eq:sigk} and \eqref{Eq:pik}, 
correspond 
to the uniform scalar and pseudoscalar modes. 
Then the chiral angle in each configuration is 
obtained by these auxiliary field modes as $\alpha=\mathrm{arctan}(\pi_0/\sigma_0)$.
We fix the chiral angle in each configuration by the chiral transformation 
so as to set $\pi_0 =0$,
and obtain quantities by using the transformed fields.
We observe finite chiral condensate in the Nambu-Goldstone phase 
and susceptibility peak at the transition
in a straightforward manner.

AFMC has several other advantages as follows.
First,
the chiral symmetry is manifest in the effective action.
Auxiliary fields are introduced as chiral partners, $\sigma_k$ and $\pi_k$,
so the chiral symmetry is obviously maintained.
Second, we can directly evaluate the fluctuation effects 
by comparing the AFMC and mean field results.
Many of the previous works on
QCD phase diagram at strong coupling are carried out 
 in the mean field analyses~\cite{MF-SCL,NLOPD,NNLO}.
\com{Next}, 
AFMC is a natural extension of the mean field treatment, \com{so}
%The auxiliary fields are assumed to be static and constant
%in the mean field treatment, and are integrated out numerically in AFMC.
 it is straightforward
to include finite coupling effects in AFMC.
Finite coupling effects have been investigated in the mean field method
~\cite{NLOPD,NNLO},
which can be extended to include auxiliary field fluctuations
in the framework of AFMC.
Finally, we can invoke various ideas to suppress the sign problem
in AFMC. 
AFMC is a generic integral technique 
and utilized in many fields, where many ideas have been proposed. 
For example, it is promising to apply 
the shifted contour formulation~\cite{avoidsign}
or the integral over Lefschetz thimbles~\cite{thimble}
to the QCD phase diagram at strong coupling in the AFMC method.

This paper is organized as follows.
In Sec.~\ref{sec:AFMC}, we explain the formulation of AFMC
in SC-LQCD.
In Sec.~\ref{sec:Results}, we show the numerical results 
on the order parameters, phase diagram, and the average phase factor.
In Sec.~\ref{sec:Discussion}, 
we discuss the order of the phase transition via the finite size scaling
\com{of the chiral susceptibility},
and numerically confirm a source of the sign problem in AFMC .
%based on the volume dependence
In Sec.~\ref{sec:SD}, we devote ourselves to a summary 
and discussion. 

\section{Auxiliary field Monte-Carlo method}
\label{sec:AFMC}
\subsection{Lattice action}
\label{subsec:Latac}
We here consider the lattice QCD 
with one species of unrooted staggered fermion
for color $\mathit{SU}(N_c)$ 
in the anisotropic Euclidean spacetime.
Throughout this paper, we work in the lattice unit $a=1$,
where $a$ is the spatial lattice spacing, 
and the case of color $\mathrm{SU}(N_c=3)$
in 3+1 dimension $(d=3)$ spacetime.
Temporal and spatial lattice sizes are denoted as $N_\tau$ and $L$,
respectively.

The partition function and action are given as,
%******************************************************************
%**********************************************************
\begin{align}
  {\cal Z}_{\mathrm{LQCD}} 
  =&  \int \Fint \left[ \chi,\chibar,U_\nu \right] e^{-S_\mathrm{LQCD}}
  \ ,\\
  S_\mathrm{LQCD}=&S_F+S_G
  \ ,\\
  S_F
  =&\frac12 \sum_x \left[
  %	 e^{\mu/\gamma^2} \bar{\chi}_x U_{0,x} \chi_{x+\hat{0}}
  %	-e^{-\mu/\gamma^2} \bar{\chi}_{x+\hat{0}} U^\dagger_{0,x} \chi_{x}
  	V^{+}_x - V^{-}_x
  		\right]
%\nonumber\\
  +\frac{1}{2} \sum_{x}\sum_{j=1}^{d} \eta_{j,x}\left[
  		 \bar{\chi}_x U_{j,x} \chi_{x+\hat{j}}
  		-\bar{\chi}_{x+\hat{j}} U^\dagger_{j,x} \chi_{x}
  		\right]
%\nonumber\\
  +m_0 \sum_{x} M_x
\label{Eq:LQCD}
\ ,\\
V^{+}_x=&\gamma e^{\mu/f(\gamma)} \bar{\chi}_x U_{0,x} \chi_{x+\hat{0}}
\ ,\\
V^{-}_x=&\gamma e^{-\mu/f(\gamma)} \bar{\chi}_{x+\hat{0}} U^\dagger_{0,x} \chi_x
\ ,\\
M_x=&\bar{\chi}_x \chi_x
\ ,\\
S_G
=&
\frac {2N_c \xi}{g_{\mrm{\tau}}^2(g_0,\xi)} \mathcal{P}_\tau 
+ \frac {2N_c}{g_{\mrm{s}}^2(g_0,\xi) \xi} \mathcal{P}_s 
\ ,\\
\mathcal{P}_i
=&
  \sum_{P_i} \left[
 	1 - \displaystyle \frac{1}{2N_c}  \Tr \left( U_{P_i} + U_{P_i}^\dagger \right) \right] 
\ (i=\tau,s)
\ ,
\end{align}
where
$\chi_x$, $U_{\nu,x}$, $U_{P_\tau}$ and $U_{P_s}$
represent the quark field, the link variable, %the lattice spacing ratio,
and the temporal and spatial plaquettes,
respectively.
$\eta_{j,x}=(-1)^{x_0+\cdots+x_{j-1}}$ is the staggered sign factor,
and $V^{\pm}_x$ and $M_x$ are mesonic composites.
%$\eta_{j,x}=(-1)^{x_0+\cdots+x_{j-1}}$ is the staggered sign factor,
Quark chemical potential $\mu$ is introduced
in the form of the temporal component of vector potential.
The physical lattice spacing ratio is introduced as
$f(\gamma)=a_\mrm{s}^\mrm{phys}/a_\tau^\mrm{phys}$.

The lattice anisotropy parameters, $\gamma$ and $\xi$, are introduced
as modification factors
of the temporal hopping term of quarks
and the temporal and spatial plaquette action terms\com{, respectively}.
Temporal and spatial plaquette couplings should satisfy 
the hypercube symmetry condition in the isotropic limit ($\xi \to 1$),
$g_{\mrm{\tau}}(g_0,1)=g_{\mrm{s}}(g_0,1)=g_{0}$.
In the continuum limit ($a \to 0$ and $g_0 \to 0$),
two anisotropy parameters should correspond 
to the physical lattice spacing ratio,
$f(\gamma)=\gamma=\xi$,
when we construct lattice QCD action requiring $a_{\mrm{s}}^{\mrm{phys}}/a_{\tau}^{\mrm{phys}}= \gamma$ in the continuum region, 
then we can define temperature as $T=f(\gamma)/N_{\tau}=\gamma/N_\tau$.
%%%%%%%%%%%%%%%%%
By comparison,
it seems more reasonable
to define  $T=\gamma^2/N_{\tau}$
due to quantum corrections in the strong coupling limit (SCL)
as discussed based on the mean field results
in Refs.~\cite{Bilic,BilicDeme,PhDFromm}.
We follow this argument and adopt $f(\gamma)=\gamma^2$.
\com{
The behavior of the chiral susceptibility suggests that 
this prescription
is reasonable even with fluctuations
as shown later in \secref{sec:Discussion}.
We briefly summarize the mean field arguments given
in Refs.~\cite{Bilic,BilicDeme,PhDFromm}
in Appendix \ref{App:fgamma}.
}

In SCL, we can ignore the plaquette action terms $S_G$,
which are proportional to $1/g^2$.
The above lattice QCD action in the chiral limit $m_0\to 0$ 
has chiral symmetry $U(1)_V \times U(1)_A$. 

\subsection{Effective action}
\label{subsec:Formalism}
In the present formulation, we have four main steps
to obtain physical observables.
First, we integrate out the lattice partition function
over spatial link variables in the strong-coupling limit.
Second,
we introduce the auxiliary fields for the mesonic composites
and convert the four-Fermi interaction terms to the fermion bilinear form.
Third, 
we perform the integral over the fermion fields and temporal link variables
analytically, 
and obtain the effective action of the auxiliary fields.
Finally, we carry out the Monte-Carlo integral over the auxiliary fields.

In the first step,
we obtain the SCL effective action by integrating out spatial link 
variables~\cite{SCL,large_d,MF-SCL,Faldt,BilicDeme,Bilic,NLOPD,NNLO,Jolicoeur},
\begin{align}
S_\mathrm{eff}
&=\frac12 \sum_x \left[ V^{+}_x - V^{-}_x \right]
- \displaystyle \frac {1}{4N_c} \sum_{x,j} M_x M_{x+\hat{j}}
+m_0 \sum_{x} M_x
\label{Eq:Seff}
\ .
\end{align}
Here we adopt the effective action in the leading order of
the $1/d$ expansion~\cite{large_d},
where $d$ is the spatial dimension, $d=3$.

The large dimensional expansion ($1/d$ expansion) is a scheme
to truncate the interaction terms systematically.
We assume that the quark fields scale as $d^{-1/4}$,
then the mesonic hopping terms, second terms in Eq.~\eqref{Eq:Seff},
stay finite at large $d$. For color SU(3), spatial link integral
also gives rise to spatial baryonic hopping terms which contain six quarks 
and sum over spatial directions, and is proportional 
to $\mathcal{O}(1/\sqrt{d})$.
In the leading order of the $1/d$ expansion adopted here,
we do not include this spatial baryonic hopping terms. 
One may suspect that ignoring the spatial baryonic hopping
corresponds to replacing color SU(3) link integral with color U(3)
and that we cannot take account of baryonic effects,
but this is not true.
Baryon effects arise\sout{s} from the temporal hopping term of quarks,
first term\comm{s} in Eq.~\eqref{Eq:Seff}.
As we discuss later, temporal link integral is carried out
exactly under the periodic and anti-periodic boundary conditions
in the temporal direction for link variables and quarks, respectively.
Baryon contribution naturally appears from the cubic terms 
of the temporal link variables, 
$(\bar{\chi}_{x} U_{0,x} \chi_{ x+\hat{0} })^3$.

In the second step, we transform the four-Fermi interactions,
the second terms in Eq.~\eqref{Eq:Seff},
to the fermion-bilinear form.
By using the Fourier transformation in spatial 
\com{coordinates}
$M_{x=(\bold{x},\tau)}=\sum_{\bold{k}} e^{i\bold{k}\cdot\bold{x}} M_{\bold{k},\tau}$,
the interaction terms read
\begin{align}
%S_s
%&\equiv
- \displaystyle \frac {1}{4N_c} \sum_{x,j} M_x M_{x+\hat{j}} %\nonumber \\
%\ , \\
&= -\frac{L^3}{4N_c}
	\sum_{\bold{k}, \tau}
	f(\bold{k})\, M_{-\bold{k},\tau}\, M_{\bold{k},\tau}
\ \nonumber\\
&=- \frac {L^3}{4N_c} \sum_{\bk, \tau, f(\bold{k})>0} f(\bold{k}) 
(M_{\bk,\tau} M_{-\bk,\tau}
%+ \displaystyle \frac {L^3}{4N_c} \sum_{\bk, \tau, f(\bold{k})>0} f(\bold{k}) 
-M_{\bar{\bk},\tau} M_{-\bar{\bk},\tau})
\ ,
\label{Eq:SsFT}
\end{align}
where 
$f(\bold{k})=\sum_j \cos\, k_j$ and $\bar{\bk}=\bk+(\pi,\pi,\pi)$.
For later use,
we divide the momentum region into 
the positive ($f(\bold{k})>0$) and negative ($f(\bold{k})<0$) modes. 
In \com{the} last line of Eq.~(\ref{Eq:SsFT}), we use  
the relation $f(\bar{\bold{k}})=-f(\bold{k})$.

We introduce the auxiliary fields via the extended Hubbard-Stratonovich (EHS) transformation~\cite{NLOPD,NNLO}.
We can bosonize any kind of composite product
by introducing two auxiliary fields simultaneously,
\begin{align}
e^{\alpha A B}  
&= \int\, d\varphi\, d\phi\,
	e^{-\alpha\left\{
        \left[
		 \varphi-(A+B)/2
        \right]^2
        +\left[
		 \phi - i(A-B)/2
        \right]^2
		\right\}+ \alpha AB}
\nonumber\\
&= \int\, d\varphi\, d\phi 
	e^{-\alpha\left\{
		\varphi^2-(A+B)\varphi
		+ \phi^2 - i(A-B)\phi
		\right\}}
\nn \\
&= \int\, d\psi\, d\psi^*\,
	e^{-\alpha\left\{
	\psi^* \psi-{A}\psi-\psi^* B
	\right\}}
\ ,\label{Eq:EHSp}
\end{align}
where $\psi=\varphi +i \phi$ and $d\psi\,d\psi^*=d\mathrm{Re}\psi\,d\mathrm{Im}\psi=d\varphi d\phi$.
When the two composites are the same, $A=B$,
\equref{Eq:EHSp} corresponds to the bosonization 
of attractive interaction terms.
For the bosonization of interaction terms which lead to repulsive potential in the mean-field approximation,
we need to introduce complex number coefficients,
\begin{align}
e^{-\alpha A B}  
&= \int\, d\psi\, d\psi^*\,
	e^{-\alpha\left\{
	\psi^*\psi-i{A}\psi-i\psi^* B
	\right\}}
\label{Eq:EHSm}
\ .
\end{align}
The bosonization of the interaction terms in Eq.~\eqref{Eq:SsFT}
is carried out as
%%%%%%%%%%%%%%%%%%
\begin{align}
&\exp\Biggl\{\sum_{\vkt,f(\vk)>0} \alpha f(\bold{k})
M_\vkmt M_\vkt\Biggr\}
\nonumber\\
&=\int
\mathcal{D}[\sigma]\,
%d\sigma_{\vkt}\,d\sigma_{\vkt}^*\,
\exp\Bigl\{
-\sum_{\vkt,f(\vk)>0} \alpha f(\vk)
%\nonumber\\
%&\hspace*{1.5cm}
\times
\bigl[ |\sigma_{\vkt}|^2
+\sigma^*_\vkt M_\vkt
+M_\vkmt \sigma_\vkt 
\bigr]
\Bigr\}
\nonumber \\
&=\int  \mathcal{D}[\sigma]\,
%d\sigma_{\vkt}\,d\sigma_{\vkt}^*\,
  \exp \Bigl\{ 
    -\sum_{\vkt,f(\vk)>0} \alpha f(\vk) |\sigma_{\vkt}|^2
%\nonumber \\
%& \hspace*{2.0cm}
    -\frac{1}{4N_c}\sum_{x,j} \left[ \sigma_{x+\hat{j}} +\sigma_{x-\hat{j}} \right] M_x
  \Bigr\}
\ ,\\
%\end{align}
%%%%%%%%%%%%%%%%%%%
%%%%%%%%%%%%%%%%%%%
%\begin{align}
&\exp\left\{
-\sum_{\vkt,f(\vk)>0}
\alpha f(\bold{k})
M_\vkmbt M_\vkbt\right\}
\nonumber\\
&=\int 
\mathcal{D}[\pi]\,
%d\pi_{\vkt}\,d\pi_{\vkt}^*\,
\exp\Bigl\{
-\sum_{\vkt,f(\vk)>0} \alpha f(\vk)
%\nonumber\\
%&\hspace*{1.5cm}
\times
\left[ |\pi_{\vkt}|^2 -i\pi^*_\vkt M_\vkbt -iM_\vkmbt \pi_\vkt 
\right]
\Bigr\}
\nonumber \\
&=\int \mathcal{D}[\pi]\,
  \exp  \Bigl\{ 
	-\sum_{\vkt,f(\vk)>0} \alpha f(\bold{k})|\pi_{\vkt}|^2
%\nonumber \\
%&\hspace*{1.5cm}
%&\hspace*{1.0cm}
    -\frac{1}{4N_c}\sum_{x,j} \left[ (i \varepsilon \pi)_{x+\hat{j}} + (i \varepsilon \pi)_{x-\hat{j}} \right]M_x
  \Bigr\}
\ ,\\
&\sigma_x
= \sum_{\vk,f(\vk)>0} e^{i\bold{k}\cdot\bold{x}}\sigma_\vkt
\ ,\quad
\pi_x
= \sum_{\vk,f(\vk)>0} (-1)^\tau e^{i\bold{k}\cdot\bold{x}}\pi_\vkt
\label{Eq:sigma-pi_x}
\ .
\end{align}
%%%%%%%%%%%%%%%%%%
%
where 
$\mathcal{D}[\sigma]=\prod_{\vkt,f(\vk)>0} d\sigma_{\vkt}\,d\sigma_{\vkt}^*$,
$\mathcal{D}[\pi]=\prod_{\vkt,f(\vk)>0} d\pi_{\vkt}\,d\pi_{\vkt}^*$,
and
$\alpha=L^3/4N_c$.
%%%%%%%%%%%%%%%%%
\com{
The sign factor, $\varepsilon_x=(-1)^{x_0+x_1+x_2+x_3}$, corresponds to 
$\Gamma_{55}=\gamma_5\otimes\gamma_5$ in the spinor-taste space.
}
%%%%%%%%%%%%%%%%%
We introduce $\sigma_{\vk,\tau}$ and $\pi_{\vk,\tau}$ as the auxiliary fields of 
$M_\vkt$ and $iM_\vkmbt$, respectively. 
$\sigma_{\vkt}$ ($\pi_{\vkt}$) includes 
the scalar (pseudoscalar) and some parts of higher spin modes.
By construction, $\sigma_{\vkt}$ and $\pi_\vkt$ satisfy the relation
$\sigma_\vkmt=\sigma_\vkt^*$ and $\pi_\vkmt=\pi_\vkt^*$, 
which means that $\sigma_x,\ \pi_x \in \mathbb{R}$.

The bosonized effective action is given as
\begin{align}
S_\mathrm{eff}^\mathrm{EHS}
=&\frac{1}{2}\sum_x\left[V_x^+ - V_x^-\right]
 +\sum_x m_x M_x
%\nonumber\\
+\frac{L^3}{4N_c} \sum_{\vkt, f(\vk)>0}
f(\vk)\left[\left|\sigma_\vkt\right|^2+\left|\pi_\vkt\right|^2\right]
\label{Eq:SeffEHS}
\ ,
\\
m_x
=&
 m_0
 +\frac{1}{4N_c} \sum_{j}
	\left[
	 (\sigma+i\varepsilon\pi)_{x+\hat{j}}
	+(\sigma+i\varepsilon\pi)_{x-\hat{j}}
	\right]
\ .\label{Eq:meff}\\
\end{align}
%where $\varepsilon_x=(-1)^{x_0+x_1+x_2+x_3}$ corresponds to 
%$\Gamma_{55}=\gamma_5\otimes\gamma_5$ in the spinor-taste space.
The lattice QCD action Eq.~\eqref{Eq:LQCD} is invariant
under the chiral U(1) transformation,
$\chi_x \to e^{i\varepsilon_x\alpha /2 }\chi_x$,
which
%and the chiral transformation 
mixes $\sigma_\vkt$ and $\pi_\vkt$.
Thus $\sigma_\vkt$ and $\pi_\vkt$ at small $\vk$ are regarded as
the usual chiral ($\sigma$) and Nambu-Goldstone ($\pi$) fields, respectively.

In the third step,
we carry out the Grassmann and temporal link ($U_0$) integrals
analytically \cite{Faldt,BilicDeme,Bilic}.
%as shown in Appendix \ref{App:Int}.
We find the partition function and the effective action as,
\begin{align}
{\cal Z}_\mathrm{AF}%_{\mathrm{LQCD}} 
%\simeq
=& \int \Fint[\sigma_{\vkt}, \pi_{\vkt}]~e^{-S_\mathrm{eff}^\mathrm{AF}}
\ , \\
S_\mathrm{eff}^\mathrm{AF}
=&\sum_{\vkt, f(\vk)>0}
  \frac{L^3f(\vk)}{4N_c}
  \left[\left|\sigma_\vkt\right|^2+\left|\pi_\vkt\right|^2\right]
   \nonumber \\ 
-&\sum_{\bold{x}} \log\left[
	X_{N_\tau}(\bold{x})^3-2X_{N_\tau}(\bold{x})
	+2\cosh(3 N_\tau\mu/\gamma^2)\right]
\ ,
\label{Eq:SeffAF}
\end{align}
where 
we note that 
$\Fint \left[ \sigma_{\vkt}, \pi_{\vkt} \right]=\prod_{\vkt,f(\vk)>0} d \sigma_{\vkt} d\sigma_{\vkt}^{\ast} d\pi_{\vkt} d\pi_{\vkt}^{\ast}$.
$X_{N_\tau}(\bold{x})$ is a known function of 
$m_x$ and can be obtained by using a recursion formula~\cite{Faldt,Bilic,BilicDeme},
as summarized in Appendix \ref{sec:rec}.
When $m_{x=(\bold{x},\tau)}$ is 
independent of $\tau$ (static),
we obtain $X_{N_\tau}=2\cosh(N_\tau\ \mathrm{arcsinh}\ (m_x/\gamma))$. 
%\com{[T.I. $m_x \rightarrow  \sqrt{ (m_0 + b_\sigma \sigma_x)^2 + (b_\sigma \pi_x)^2}$, where $b_\sigma = d/2N_c$ ?]}

In the last step, we carry out 
%auxiliary field Monte-Carlo (AFMC) 
AFMC integral~\cite{Ohnishi:2012yb,Ichihara:2013pc}.
We numerically integrate out the auxiliary fields 
$(\sigma_\vkt, \pi_\vkt)$
based on the auxiliary field effective action, Eq.~\eqref{Eq:SeffAF},
by using the Monte-Carlo method,
%We call this treatment an auxiliary field Monte-Carlo (AFMC) method.
%
then we could take auxiliary field fluctuation effects into account.

When we perform integration, we have a sign problem in AFMC \cite{Ohnishi:2012yb,Ichihara:2013pc}.
The effective action $S_\mathrm{eff}^\mathrm{AF}$ in Eq.~\eqref{Eq:SeffAF}
contains the complex terms $X_{\Nt}$ 
via the spatial diagonal parts of the fermion matrix $I_x =2 m_{x}/\gamma$.
Auxiliary fields are real in the spacetime representation,
$\sigma_x, \pi_x \in \mathbb{R}$,
but the negative auxiliary field modes appear with imaginary coefficients
as $i\varepsilon_x \pi_x$, which come from the EHS transformation.
The imaginary part of the effective action 
gives rise to a complex phase in the statistical weight 
$\exp(-S_\mathrm{eff}^\mathrm{AF})$,
and leads to the weight cancellation.

It should be noted that the weight cancellation is weakened
in part by the phase cancellation mechanism
in low momentum auxiliary field modes.
In AFMC, 
the fermion determinant is decomposed into the one at each spatial site.
Since negative modes $\pi_\vkt$ involve $i\varepsilon_x$,
the phase on one site from low momentum $\pi_\vkt$ modes tend to be canceled 
by the phase on the nearest neighbor site.
Thus we could expect that the weight cancellation is not severe 
when low momentum modes mainly contribute.
By comparison, strong weight cancellation might arise 
from high momentum modes.
We discuss the contributions from high momentum modes
in \secref{subsec:Highmode}. 

While we have the sign problem in AFMC, 
we anticipate that we could study the QCD phase diagram 
since the long wave modes are more relevant to phase transition phenomena.
We show the results of the QCD phase transition phenomena based on AFMC 
in the next section, \secref{sec:Results}.

\section{QCD phase diagram in AFMC}
\label{sec:Results}

We show numerical results in the chiral limit $(m_0=0)$ 
on $4^3\times 4$, $6^3\times 4$, $6^3\times6$ and $8^{3}\times 8$ lattices.
We have generated the auxiliary field configurations
at several temperatures on fixed fugacity (fixed $\mu /T$) lines.
We here assume that temperature is given as $T=\gamma^2/N_\tau$ \cite{Bilic}. 
Statistical errors are evaluated in the jack-knife method;
we consider an error to be the saturated value 
after the autocorrelation disappears as shown later in \figref{Fig:JK}.

\subsection{Chiral Angle Fixing}
\label{Sec:CAF}
\begin{figure}[tbp]
 \begin{center}
  \includegraphics[width=120mm]{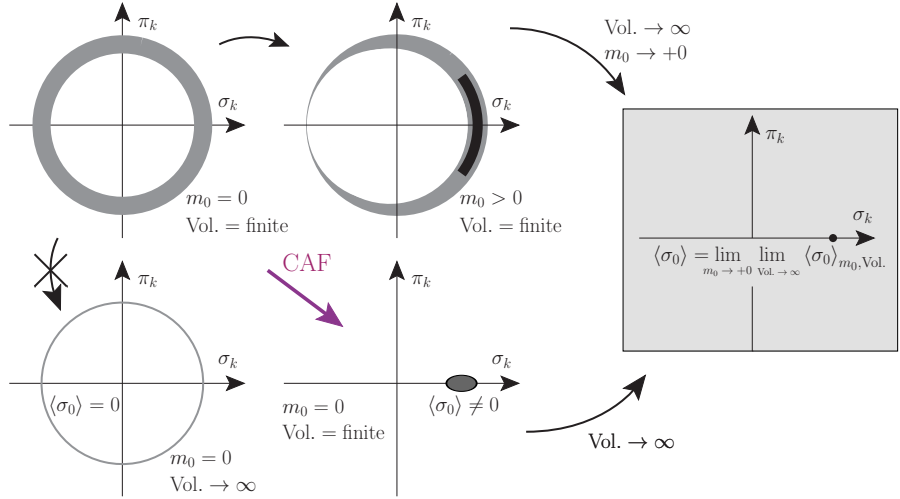}
  \end{center}
 \caption{
 Schematic picture of CAF method.
In order to obtain the chiral condensate rigorously,
we need to put a finite mass, first take thermodynamic limit 
and finally take the chiral (massless) limit
as shown in the upper panels.
In CAF, we take chiral rotation to make the $\pi_{0}$ field 
vanish, and we get the finite chiral condensate 
(center bottom panel), 
which would be close to the correct value.
}
  \label{Fig:CAF}
\end{figure}

It is a non-trivial problem how to describe the spontaneous symmetry breaking 
in Monte-Carlo calculations on a fixed finite size lattice:
the expectation value of the order parameter generally vanishes
since the distribution is symmetric under the transformation.
Rigorously, we need to take the thermodynamic limit 
with explicit symmetry breaking term,
and %we need 
to take the limit of the vanishing explicit breaking term,
as schematically shown in \figref{Fig:CAF} 
in the case of chiral symmetry.
Another method is measuring correlations of the order parameter and finite size scaling of the correlation function \cite{PhDFromm}.
These procedures are time consuming and not easy to carry out
when we have the sign problem.

%%%%%%%%%%%%%%%%%%%%%%%%%%%%%%%%%%%%%%%%%%%%%%%%%%%%%%%%%%%%%%%%%%%%%%%%%%%%%%%%
%%%%%%%%%%%%%%%%%%%%%%%%%%%%%%%%%%%%%%%%%%%%%%%%%%%%%%%%%%%%%%%%%%%%%%%%%%%%%%%%

We here propose a chiral angle fixing (CAF) method
as a prescription to calculate the chiral condensate on a fixed finite size lattice.
The effective action Eq.~\eqref{Eq:Seff} is invariant
under the chiral transformation, 
\begin{align}
\chi_x \to \chi_x' = e^{i\varepsilon_x \alpha/2} \chi_x
\ ,\quad
\chibar_x \to \chibar_x' = e^{i\varepsilon_x \alpha/2} \chibar_x\ .
\end{align}
The chiral symmetry is kept in the bosonized effective action
by introducing the chiral U(1) transformation for auxiliary fields as,
\begin{align}
\begin{pmatrix}
\sigma_k \\ \pi_k
\end{pmatrix}
\to
\begin{pmatrix}
\sigma_k' \\ \pi_k'
\end{pmatrix}
=
\begin{pmatrix}
\cos\alpha & -\sin\alpha\\
\sin\alpha& \cos\alpha
\end{pmatrix}
\begin{pmatrix}
\sigma_k \\ \pi_k
\end{pmatrix}
\ ,
\label{Eq:chitransAF}
\end{align}
where $(\sigma_k, \pi_k)$ are the temporal Fourier transform of
$(\sigma_{\vkt}, \pi_\vkt)$,
\begin{align}
\sigma_{k=(\vk,\omega)}
=& \frac{1}{N_\tau}\sum_{\tau} e^{-i\omega\tau} \sigma_{\vk,\tau}
\ , \label{Eq:sigk}\\
\pi_{k=(\vk,\omega)}
=& \frac{1}{N_\tau}\sum_{\tau} (-1)^\tau e^{-i\omega\tau} \pi_{\vk,\tau}
%\nonumber \\
%=& 
=\frac{1}{\Nt L^3} \sum_x e^{-ik \cdot x} \pi_x 
\ . \label{Eq:pik}
\end{align}
Because of the chiral symmetry,
the chiral condensate $\VEV{\sigma_0}$ vanishes
as long as the auxiliary field configurations are taken to be chiral symmetric,
as explicitly shown in Appendix~\ref{App:CAF}.

In order to avoid the vanishing chiral condensate,
we here utilize CAF.
We rotate $\sigma_0$ and $\pi_0$ modes toward the positive $\sigma_0$ direction
as schematically shown in \figref{Fig:CAF}.
All the other fields are rotated with the same angle,
$-\alpha=-\arctan (\pi_{0}/\sigma_{0})$, in each Monte-Carlo configuration.
We use these new fields to 
obtain order parameters, susceptibilities, and other quantities,
and eventually obtain finite chiral condensate.
Chiral condensate obtained in CAF should mimic the spontaneously broken
chiral condensate in the thermodynamic limit.
%%%%%%%%%%%%%
Similar prescriptions are adopted in other field of physics.
For example, we take a root mean square order parameter 
to obtain the appropriate value in spin systems~\cite{MCsim1}.
%%%%%%%%%%%%%

\subsection{Sampling and Errors}

%%%%%%%%%%%%%%%%%%%%%%%%%%%%%%%%%%%%%%%%%%%%%%%%%%%%%%%%%%%%%%%%%%%%%%%%%%%%%%%%
\begin{figure}[tb]
\centerline{\PSfig{9cm}{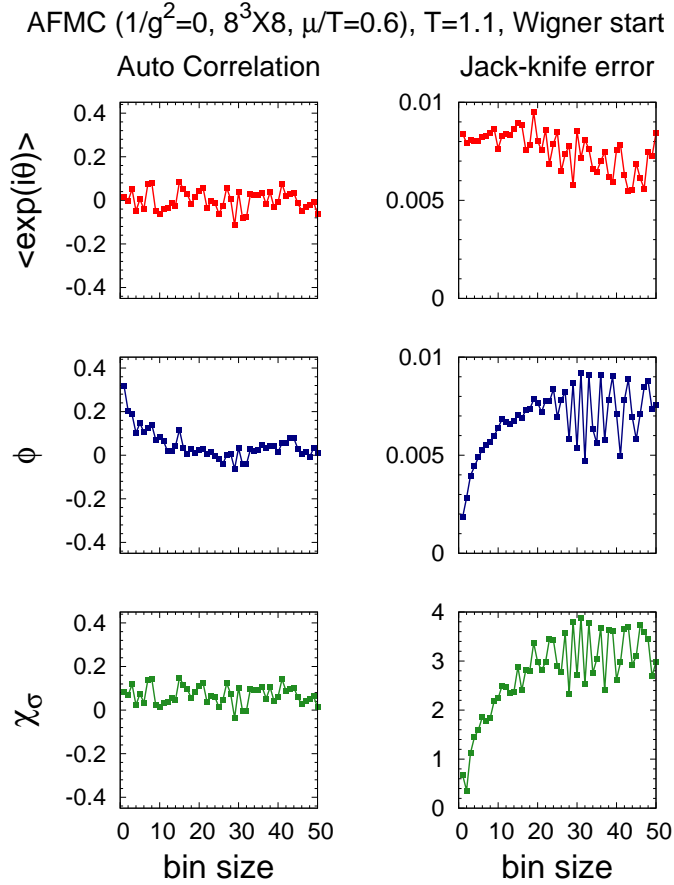}}
\caption{Autocorrelation (left column) and jack-knife error (right column) of average phase factor (top row), chiral condensate (middle row) and chiral susceptibility (bottom row) as a function of bin size 
for $\mu /T =0.6, \ T=1.1$ on a $8^3 \times 8$ lattice starting from the Wigner phase initial condition.
We adopt the saturated values after autocorrelations disappear
as the errors of calculated quantities.
}
\label{Fig:JK}
\end{figure}
%%%%%%%%%%%%%%%%%%%%%%%%%%%%%%%%%%%%%%%%%%%%%%%%%%%%%%%%%%%%%%%%%%%%%%%%%%%%%%%%

We generate auxiliary field configurations
by using the Metropolis sampling method.
We generate Markov chains
starting from two types of initial conditions:
the Wigner phase ($\sigma_x =0.01,\ \pi_x =0$)
and the Nambu-Goldstone (NG) phase ($\sigma_x =2,\ \pi_x =0$) initial conditions.

For each $\tau$, we generate a candidate auxiliary field configuration
$(\sigma'_{\vkt},\pi'_{\vkt})$
by adding random numbers to the current configuration $(\sigma_{\vkt},\pi_{\vkt})$
for all spatial momenta $\vk$ at a time,
and judge
whether the new configuration is accepted or not\comm{ by using the 
Metropolis algorithm}.
Since it is time consuming to update each auxiliary field mode separately, 
we update all spatial momentum modes in one step
at the cost of an acceptance probability.
\comm{We have tuned the strength (standard deviation) 
of the random numbers added to \com{the current configuration}
$(\sigma_{\vkt},\pi_{\vkt})$ \com{in order} to keep the acceptance probability around 50 \%
at each $(T,\mu)$.}
It should be noted that
the acceptance probability is larger in \sout{the} the present 
$(\sigma_{\vkt},\pi_{\vkt})$ sampling procedure in each $\tau$
compared with updating 
%all energy-momentum modes 
$(\sigma_k ,\pi_k)$ in the whole momentum space at a time\comm{ as done 
in our preliminary work}~\cite{Ohnishi:2012yb}.
\comm{For example,
on a $4^3\times 4$ lattice at around the critical temperature at $\mu=0$,
we find that the acceptance probability decreases 
to $15-20$ \% in the $(\sigma_k,\pi_k)$ sampling 
from $\sim 50 \%$ in the $(\sigma_{\vkt},\pi_{\vkt})$ sampling,
when we adopt the same standard deviation.
On a larger lattice, the acceptance probability difference in the two sampling
methods would be larger.}
%as that in the $(\sigma_{\vkt},\pi_{\vkt})$ sampling.

We evaluate errors of calculated quantities in the jack-knife method \cite{JK}.
The evaluated errors of the chiral condensate $\phi$ 
are shown as a function of bin size in the right middle panel of \figref{Fig:JK}.
Since the Metropolis samples are generated sequentially
in the Markov chain, subsequent events are correlated.
This autocorrelation disappears when the Metropolis time difference is
large enough.
In the jack-knife method, we group the data into bins
and regard the set of configurations except for those in a specified bin
as a jack-knife sample.
We find that the autocorrelation disappears for the bin size larger than 30
in this case.
The jack-knife error increases with increasing bin size,
and eventually saturates.
We adopt the saturated value of the jack-knife error after the autocorrelation
disappears as the error of the calculated quantity
as in the standard jack-knife treatment.
\com{The errors are found to be small enough to discuss the phase transition.
For example we find $\Delta \phi \lesssim 0.01$ in \figref{Fig:JK},
and the value is small compared with its mean value $\sout{\Delta} \phi \simeq 0.08$ shown in \figref{Fig:OPs}.
}

\subsection{Order Parameters}

%%%%%%%%%%%%%%%%%%%%%%%%%%%%%%%%%%%%%%%%%%%%%%%%%%%%%%%%%%%%%%%%%%%%%%%%%%%%%%%%
\begin{figure}[tbh]
\begin{center}
\PSfig{7.5cm}{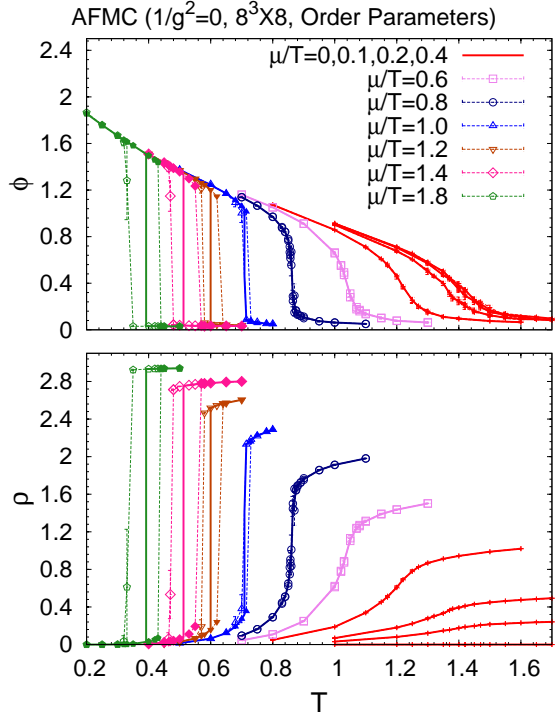}
\end{center}
\caption{Chiral condensate (upper panel)
and quark number density (lower panel)
as a function of temperature ($T$) for a fixed $\mu/T (0\leq \mu/T \leq 1.8)$
on a $8^3 \times 8$ lattice.
Open symbols, filled symbols and lines show 
the results with the Wigner and NG initial conditions
and those in the realized phase, respectively.
}
\label{Fig:OPs}
\end{figure}
%%%%%%%%%%%%%%%%%%%%%%%%%%%%%%%%%%%%%%%%%%%%%%%%%%%%%%%%%%%%%%%%%%%%%%%%%%%%%%%%

%%%%%%%%%%%%%%%%%%%%%%%%%%%%%%%%%%%%%%%%%%%%%%%%%%%%%%%%%%%%%%%%%%%%%%%%%%%%%%%%
\begin{figure}[tbh]
\begin{center}
\includegraphics[width=6.5cm,angle=-90]{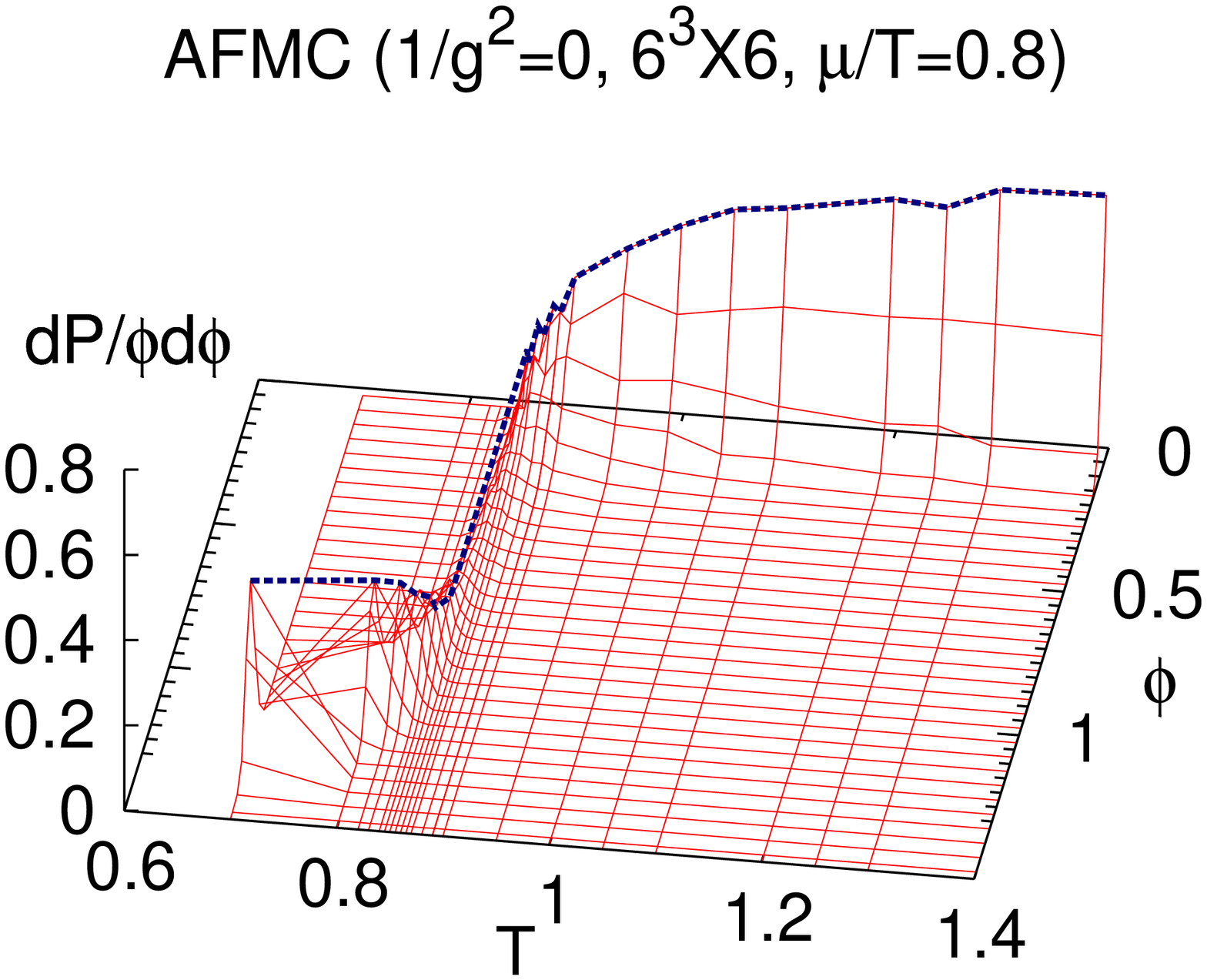}\\
\includegraphics[width=6.5cm,angle=270]{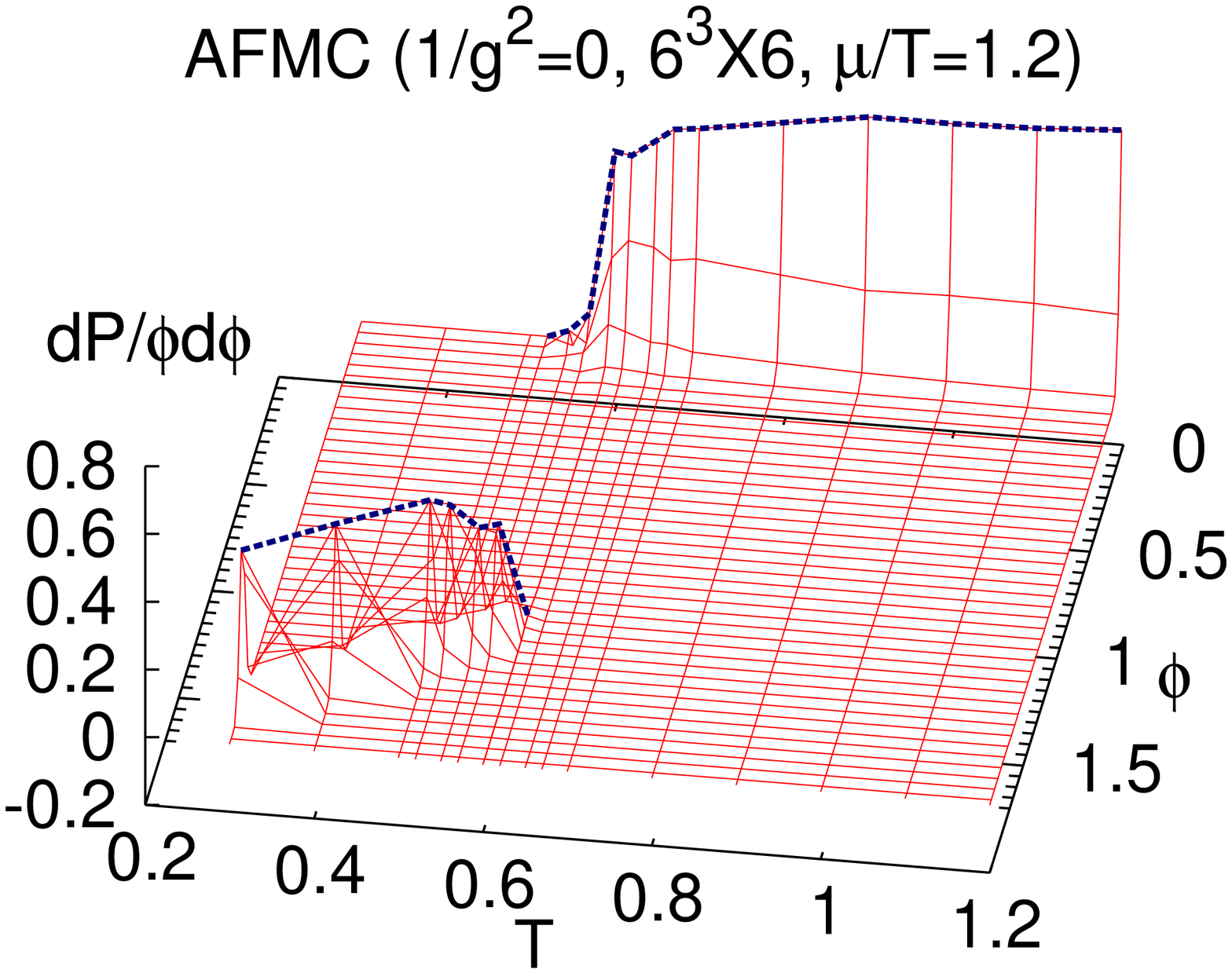}
\end{center}
\caption{
Chiral condensate ($\phi$) distribution as a function of temperature ($T$) 
and $\phi$
on a $6^3 \times 6$ lattice for $\mu/T =0.8$ (top panel) and 1.2 (bottom panel).}
\label{Fig:OPdist}
\end{figure}
%%%%%%%%%%%%%%%%%%%%%%%%%%%%%%%%%%%%%%%%%%%%%%%%%%%%%%%%%%%%%%%%%%%%%%%%%%%%%%%%

In \figref{Fig:OPs},
we show the chiral condensate, 
$\phi = \expv{\sigma_0}$,
and the quark number density $\rho_q$ after CAF,
\begin{align}
\expv{\sigma_0}
=&\frac{1}{L^3N_\tau}\,\frac{\partial \ln Z}{\partial m_0}
=-\expv{\chibar\chi}
\ , \label{Eq:defsigma}\\
\rho_q
=&\frac{T}{L^3}\,\frac{\partial \ln Z}{\partial \mu}
\ , \label{Eq:defrho}
\end{align}
as a function of temperature ($T$)
on a $8^3\times 8$ lattice.
Necessary formulae to obtain these quantities are summarized
in Appendix \ref{sec:rec}. 
We also show the distribution of $\phi$ in \figref{Fig:OPdist}.

The order parameters, $\phi$ and $\rho_q$, clearly show
the phase transition behavior.
With increasing $T$ for fixed $\mu/T$, 
the chiral condensate $\phi$ slowly decreases at low $T$,
 shows rapid or discontinuous decrease at around the transition temperature, 
and stays to be small at higher $T$.
The quark number density $\rho_q$ also shows the existence 
of phase transition at finite $\mu$.

The order of the phase transition can be deduced 
from the behavior of $\phi$, $\rho_q$ and the $\phi$ distribution 
on a small lattice~\cite{Ohnishi:2012yb,Ichihara:2013pc}.
The chiral condensate $\phi$ and the quark number density $\rho_q$ smoothly
change around the (pseudo-)critical temperature ($T_c$)
at small $\mu/T$.
Additionally, the $\phi$ distribution has a single peak
% at low $\mu /T$ 
as shown in the top panel of Fig.~\ref{Fig:OPdist}.
These observations suggest that the phase transition is crossover or the second order
at small $\mu/T$ on a large size lattice.
We refer to this $\mu/T$ region 
as the {\em would-be second order} region.

By comparison,
the order parameters show hysteresis behavior in the large $\mu/T$ region.
As shown by dashed lines in \figref{Fig:OPs}, 
two distinct results of $\phi$ and $\rho_q$ depend on the initial conditions,
the Wigner phase and the NG phase initial conditions.
The temperature of sudden $\phi$ change for the NG initial condition
is larger than that for the Wigner initial condition.
The distribution of $\phi$ shows a double peak as shown
in the bottom panel of \figref{Fig:OPdist}.
In terms of the effective potential, the dependence of initial conditions indicates that there exist two local minima, 
which are separated by a barrier.
In the hysteresis region,
the transition between the two local minima is suppressed by the barrier
and Metropolis samples stay around the local minimum close to the initial 
condition. At the temperature of sudden $\phi$ change,
the barrier height becomes small enough for the Metropolis samples 
to overcome.
These results suggest that the phase transition
is the first order at large $\mu/T$.
We refer to this $\mu/T$ region 
as the {\em would-be first order} region.

%%%%%%%%%%%%%%%%%%%%%%%%%%%%%%%%%%%%%%%%%%%%%%%%%%%%%%%%%%%%%%%%%%%%%%%%%%%%%%%%
\begin{figure}[tb]
  \begin{center}
   \includegraphics[width=80mm]{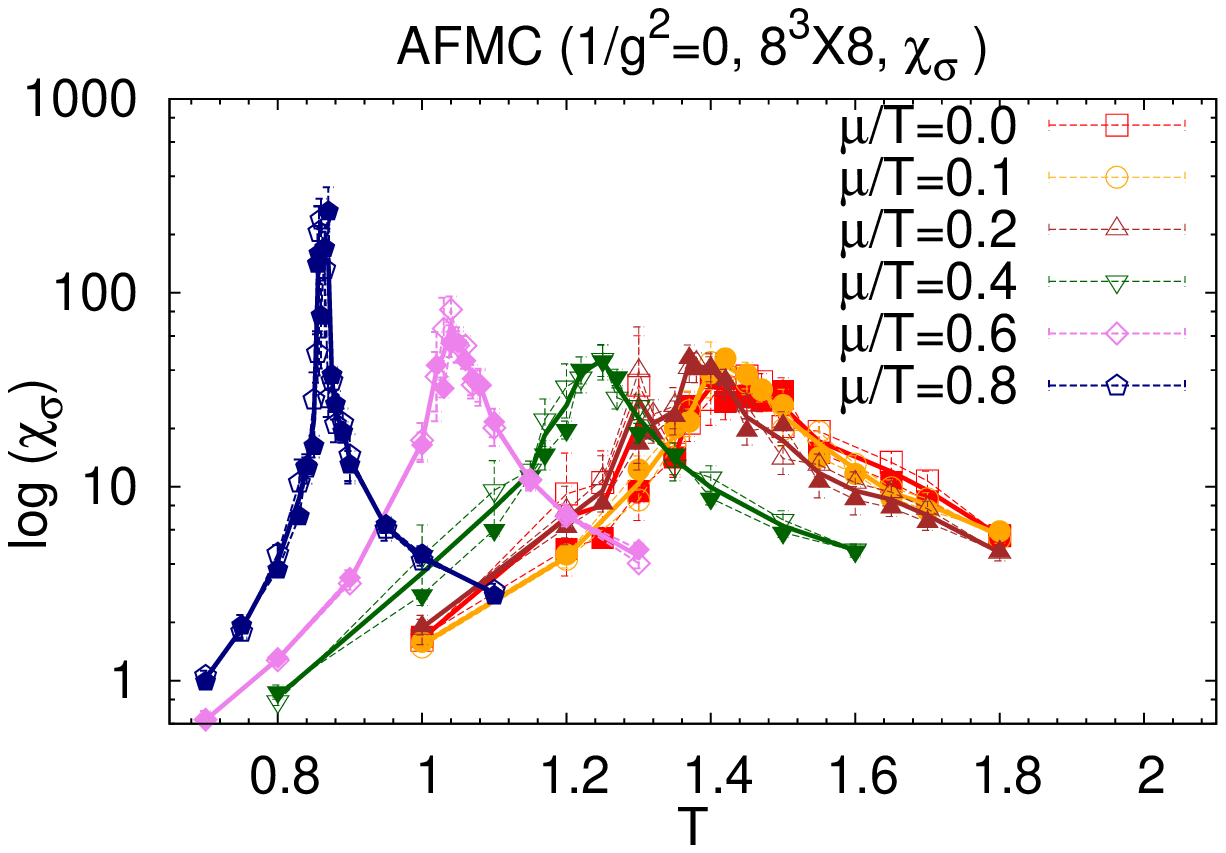}%
~\includegraphics[width=80mm]{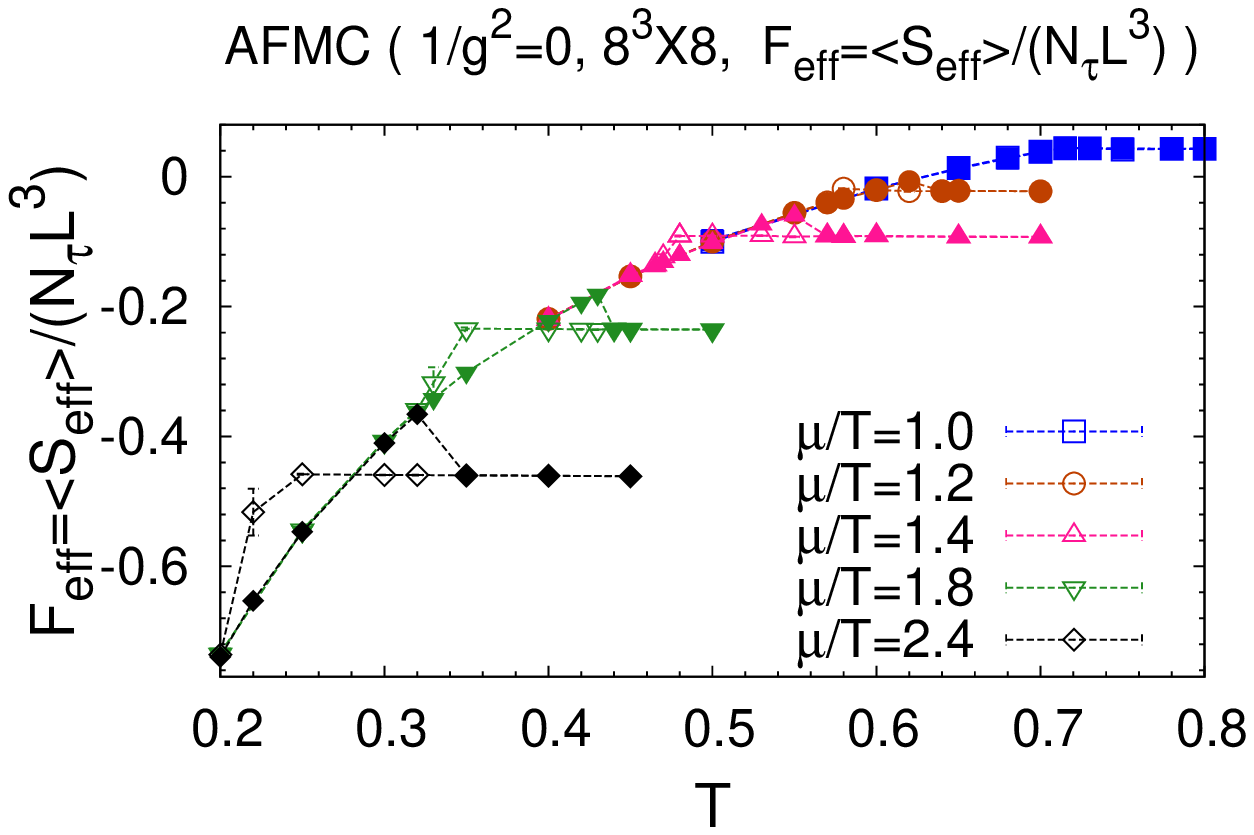}
  \end{center}
\caption{Chiral susceptibility $\chi_{\sigma}$ (left panel) and effective potential $F_{\mrm{eff}}=\expv{S_{\mrm{eff}}}/(\Nt L^3)$ (right panel) as a function of temperature on a $8^3 \times 8$ lattice. 
We determine the phase boundary 
by using $\chi_{\sigma}$ in the would-be 2nd order region
and $F_{\mrm{eff}}$ in the would-be 1st order region.
Open symbols and filled symbols show results with the Wigner and NG initial conditions,
respectively.}
\label{Fig:CF}
\end{figure}
%%%%%%%%%%%%%%%%%%%%%%%%%%%%%%%%%%%%%%%%%%%%%%%%%%%%%%%%%%%%%%%%%%%%%%%%%%%%%%%%

\subsection{Phase Diagram}

\begin{figure}[tb]
  \begin{center}
    \includegraphics[width=8cm]{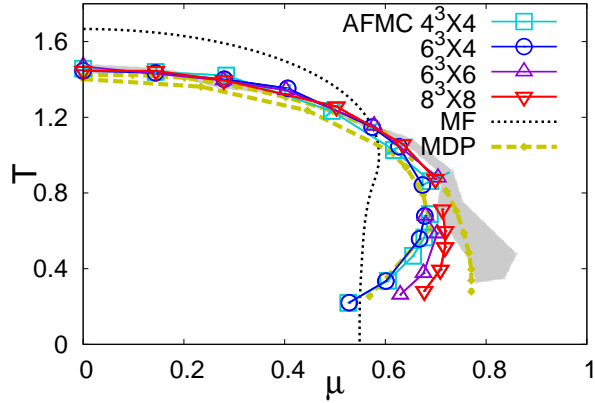}
  \end{center}
\caption{Phase diagram in AFMC on $4^{3}\times 4$, $6^{3}\times \Nt$ 
$(\Nt=4,6)$ and $8^{3}\times 8$ lattices. 
Short- and long-dashed lines show
the mean-field (MF)~\cite{MF-SCL,NLOPD,NNLO} and the monomer-dimer-polymer (MDP) results~\cite{MDP,PhDFromm},
respectively.
\com{
Two 
%different 
lines in MDP show the results for $N_\tau =4$ 
(the left line) and those extrapolated to
%\sout{for} 
$N_\tau \to \infty$ (the right line).
}
Shaded area shows the phase boundary extrapolated to $N_\tau \to \infty $ \com{in AFMC}.
}
\label{Fig:PB}
\end{figure}

We shall now discuss the QCD phase diagram in AFMC.
In \figref{Fig:PB}, we show the QCD phase diagram
% in AFMC 
%with CAF 
for various lattice sizes.
We define the (pseudo-)critical temperature $T_c$ as a peak position 
of the chiral susceptibility 
$\chi_{\sigma}$
($=\partial^2 \ln Z / \partial m_{0}^{2} / L^3 \Nt$,
see also Appendix \ref{sec:rec}) shown in \figref{Fig:CF}
in the would-be second order region.
We determine the peak position by fitting the susceptibility 
with a quadratic function. 
The errors are comprised of both statistical and systematic errors. 
We fit $\chi_\sigma$ as a function of $T$
with statistical errors obtained in the jack-knife method.
In order to evaluate the systematic error,
we change the fitting range as long as the fitted quadratic function describes 
an appropriate peak position. 
We take notice that we do not fit $\chi_{\sigma}$ as a function of $T$
in each jack-knife sample.
%%%%%%%%%%%%%%%%%

%%%%%%%%%%%%%%%%%%%%%%%%%%%%%%%%%%%%%%%%%%%%%%%%%%%%%%%%%%%%%%%%%%%%%%%%%%%%%%%%
%%%%%%%%%%%%%%%%%%%%%%%%%%%%%%%%%%%%%%%%%%%%%%%%%%%%%%%%%%%%%%%%%%%%%%%%%%%%%%%%

In the would-be first order region of $\mu/T$, 
we determine the phase boundary by comparing the expectation values
of effective action $\VEV{S_\mathrm{eff}}$
in the configurations sampled from the Wigner and NG 
phase initial conditions.
We define $T_c$ as the temperature 
where $\VEV{S_\mathrm{eff}}$ with the Wigner initial condition
becomes lower than that with the NG initial condition as shown in
\figref{Fig:CF}.
We have adopted this prescription, since it is not easy
to obtain equilibrium configurations over the two phases
when the thermodynamic potential barrier is high.
At large $\mu/T$, Metropolis samples in one sequence stay in the local minimum
around the initial condition,
and we need very large sampling steps to overcome the barrier.

In \figref{Fig:PB}, 
we compare the AFMC phase boundary
with that in the mean field approximation~\cite{MF-SCL,NLOPD,NNLO}
and in the MDP simulation~\cite{MDP,PhDFromm}
in the strong coupling limit.
Compared with the MF results, $T_c$ at low $\mu$ is found to be smaller,
and NG phase is found to be extended
in the finite $\mu$ region in both MDP~\cite{MDP,PhDFromm} and AFMC. 
As found in previous works~\cite{Ohnishi:2012yb,Ichihara:2013pc},
the phase boundary is approximately independent of the lattice size
in the would-be second order region.
The would-be first order phase boundary
is insensitive to the spatial lattice size
but is found to depend on the temporal lattice size.
With increasing temporal lattice size,
the transition chemical potential $\mu_c$ becomes larger,  
which is consistent with MDP~\cite{MDP}.
Phase boundary extrapolated to $N_\tau \to \infty$ 
is \com{obtained by assuming that the spatial size dependence is negligible.
The extrapolated boundary is shown by the shaded area,} and is found to be consistent
with the continuous time MDP results with the same limit,
$N_\tau \to \infty$ with keeping $\gamma^2/N_\tau$ finite.

Spatial lattice size independence of the phase boundary may be
understood as a consequence of almost decoupled pions.
The zero momentum pion can be absorbed into the chiral condensate
via the chiral rotation and has no effects on the transition.
Finite momentum pion modes have finite excitation energy,
then we do not have soft modes in the would-be first order 
region on a small size lattice.
For a more serious estimate of the size dependence,
we need larger lattice calculations.

We find that the would-be first order phase boundary 
has a positive slope, $d\mu/dT>0$, at low $T$.
The Clausius-Clapeyron relation reads
$d\mu/dT|_\mathrm{1st} = -(s^\mathrm{W}-s^\mathrm{NG})/(\rho_q^\mathrm{W}-\rho_q^\mathrm{NG})$,
where $s^\mathrm{W,NG}$ and $\rho_q^\mathrm{W,NG}$ are 
the entropy density and quark number density in the Wigner and NG phases,
respectively.
Since $\rho_q$ is higher in the Wigner phase as shown in \figref{Fig:OPs},
the entropy density should be smaller in the Wigner phase.
This is because $\rho_q$ is close to the saturated value, $\rho_q \sim 3 =N_c$,
in the Wigner phase, then the entropy is carried by the hole 
from the fully saturated state.
Similar behavior is found in the mean-field treatment in the strong coupling limit~\cite{MF-SCL}.
In order to avoid the quark number density saturation, which is a lattice artifact,
we may need to adopt a larger $N_\tau$~\cite{MDP}
or to take account of finite coupling effects~\cite{NLOchiral,NLOPD,Bilic,BilicDeme,Faldt,Jolicoeur,NNLO}.

%%%%%%%%%%%%%%%%%%%%%%%%%%%%%%%%%%%%%%%%%%%%%%%%%%%%%%%%%%%%%%%%%%%%%%%%%%%%%%%%
\begin{figure}[tb]
  \begin{center}
   \includegraphics[width=75mm]{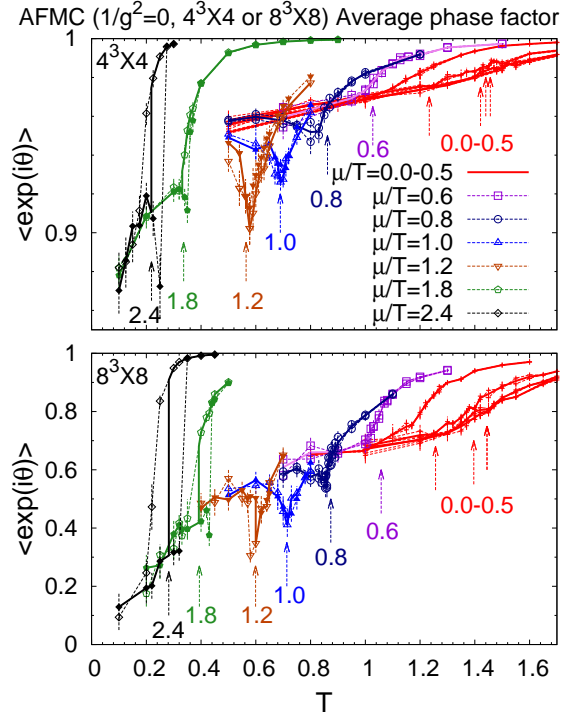}
  \end{center}
\caption{Average phase factor on $4^3 \times 4$ (top panel) and
$8^3 \times 8$ (bottom panel) lattices as a function of temperature.
Open symbols, filled symbols and lines show
the results with the Wigner and NG initial conditions
and those in the realized phase, respectively.
\com{Arrows indicate phase transition temperatures,
and the numbers under arrows represent $\mu /T$.}}
\label{Fig:sign}
\end{figure}
%%%%%%%%%%%%%%%%%%%%%%%%%%%%%%%%%%%%%%%%%%%%%%%%%%%%%%%%%%%%%%%%%%%%%%%%%%%%%%%%

%%%%%%%%%%%%%%%%%%%%%%%%%%%%%%%%%%%%%%%%%%%%%%%%%%%%%%%%%%%%%%%%%%%%%%%%%%%%%%%%
\begin{figure}[tbhp]
  \begin{center}
   \includegraphics[width=60mm,angle=270]{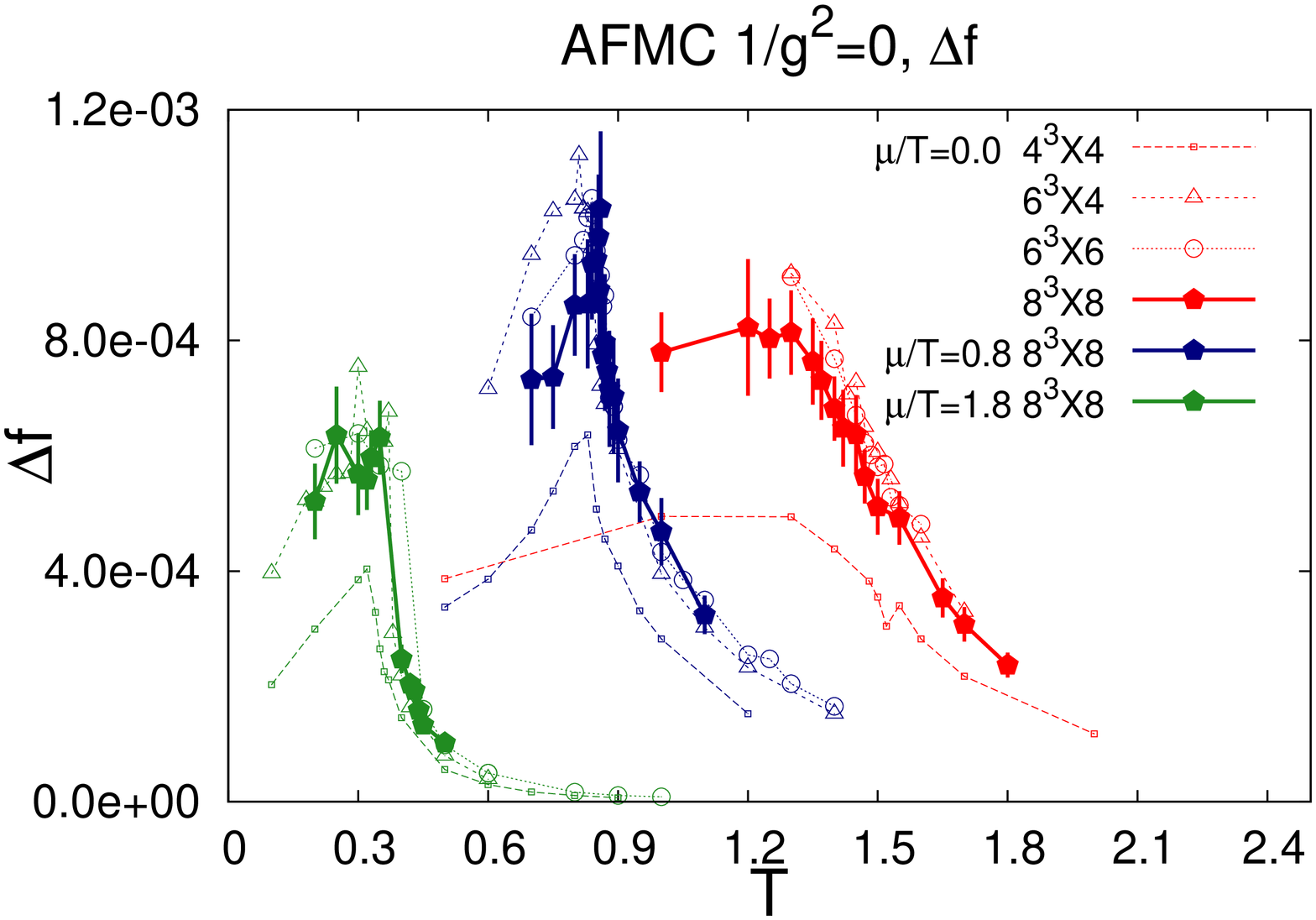}
  \end{center}
\caption{Free energy density difference $\Delta f$ 
in full and phase quenched simulations as a function of temperature
on $4^3 \times 4$, $6^3 \times 4$, $6^3 \times 6$ and $8^3 \times 8$ lattices.
We only show jack-knife errors with $8^3 \times 8$ lattice results.
}
\label{Fig:df}
\end{figure}
%%%%%%%%%%%%%%%%%%%%%%%%%%%%%%%%%%%%%%%%%%%%%%%%%%%%%%%%%%%%%%%%%%%%%%%%%%%%%%%%

\subsection{Average Phase Factor}

In \figref{Fig:sign},
we show the average phase factor $\expv{e^{\mrm{i}\theta}}$
as a function of $T$ on $8^3 \times 8$ and $4^3 \times 4$ lattices,
where $\theta$ is a complex phase of the fermion determinant
in each Monte-Carlo configuration.
The average phase factor shows the severity
of the weight cancellation;
we have almost no weight cancellation when $\expv{e^{\mrm{i}\theta}} \simeq 1$, 
and the weight cancellation is severe 
in the cases where $\expv{e^{\mrm{i}\theta}} \simeq 0$.
The average phase factor has a tendency to 
increase at large $\mu$ except for the transition region.
This trend can be understood from the effective action in \equref{Eq:SeffAF}.
The complex phase appears from $X_{\Nt}$ terms containing auxiliary fields,
and their contribution generally becomes smaller
compared with the chemical potential term, $2\cosh(3N_\tau\mu/\gamma^2 )$,
at large $\mu$.
In the phase transition region, 
fluctuation effects of the auxiliary fields are decisive
and finite momentum auxiliary fields might contribute significantly, 
which leads to a small average phase factor.

The average phase factor on a $4^3 \times 4$ lattice,
$\expv{e^{\mrm{i}\theta}} \gtrsim 0.9$,
is practically large enough to 
keep statistical precision.
%credibility.
By comparison, 
the smallest average phase factor on a $8^3 \times 8$ lattice is around 0.1
at low temperature on a $\mu/T=2.4$ line. 
Even with this average phase factor,
uncertainty of the phase boundary shown in \figref{Fig:PB}
is found to be small enough to discuss the fluctuation effects.

We show the severity of the sign problem in AFMC in \figref{Fig:df}.
The severity is characterized by the difference of the free energy density
in full and phase quenched (p.q.) MC simulations,
$\Delta f =f_{\mrm{full}} -f_{\mrm{p.q.}} $ 
which is related to the average phase factor,
$e^{-\Omega \Delta f}=\expv{e^{\mrm{i}\theta}}$,
where $\Omega =\Nt L^3$ is the spacetime volume.
While $\Delta f$ takes smaller values on a $4^3 \times 4$ lattice 
\com{than those on larger lattices},
it takes similar values on lattices with larger spatial size $L\geq 6$.
We expect that $\Delta f$ in AFMC for larger lattices
would take values similar to those on a $8^3 \times 8$ lattice.

We find that $\Delta f$ in AFMC is about twice as large as that in MDP 
when we compare the results at similar $(\mu,T)$~\cite{PhDFromm,MDPsign}.
It means that the sign problem in AFMC is more severe than that in MDP.
It is desired to develop a scheme to reduce $\Delta f$ in AFMC
on larger lattices.
In \secref{subsec:Highmode},
we search for a possible way to weaken the weight cancellation
by cutting off high momentum auxiliary fields.

\section{Discussion} \label{sec:Discussion}
\subsection{Volume Dependence of Chiral Susceptibility}
We perform a finite size scaling analysis of the chiral susceptibility
to discuss the phase transition order in the low chemical potential region. 
We expect that the phase transition is the second order at small $\mu /T$
according to the mean-field results and O(2) symmetry arguments.
The latter states that the fluctuation induced first order phase transition
is not realized as for O(2) symmetry \cite{PW84}.
%%%%%%%%%%%%%%%%%%%%%%%%%%%%%%%%%%%%%%%%%%%%%%%%%%%%%%%%%%%%%%%%%%%%%%%%%%%%%%%%
\begin{figure}[tb]
  \begin{center}
   \includegraphics[width=80mm]{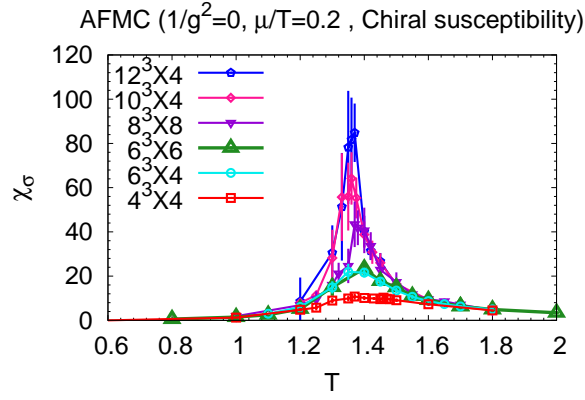}
  \end{center}
\caption{Lattice size dependence of \com{the} chiral susceptibility for $\mu /T=0.2$.
Squares, circles, triangles, upside-down triangles, diamonds, and pentagons show the results on $4^3\times 4,\ 6^3\times 4,\ 6^3 \times 6,\ 10^3 \times 4$, and $12^3 \times 4$ lattices, respectively.
The peak height depends on spatial lattice size.}
\label{Fig:chicom}
\end{figure}
%%%%%%%%%%%%%%%%%%%%%%%%%%%%%%%%%%%%%%%%%%%%%%%%%%%%%%%%%%%%%%%%%%%%%%%%%%%%%%%%

In \figref{Fig:chicom},
we show the chiral susceptibility for fixed $\mu /T=0.2$ 
on various size lattices.
In addition to $4^3 \times 4,\ 6^3 \times 4,\ 6^3 \times 6$ and $8^3 \times 8$ results,
we also show larger lattice results, $10^3 \times 4$, and $ 12^3 \times 4$.
From this comparison,
we find that $\chi_\sigma$ has a peak at the same $T$
for different lattice sizes,
and that the peak height on $6^3 \times 4$ and $6^3 \times 6$
lattices are almost the same.
These two findings suggest that it is reasonable to define the temperature
as $T=\gamma^2/N_\tau$ in the strong coupling limit.
We also find that 
the peak height of the susceptibility increases
with increasing spatial lattice size.
The divergence of the susceptibility in the thermodynamic limit
signals the first or second order phase transition.

%%%%%%%%%%%%%%%%%%%%%%%%%%%%%%%%%%%%%%%%%%%%%%%%%%%%%%%%%%%%%%%%%%%%%%%%%%%%%%%%
\begin{figure}[tb]
  \begin{center}
   \includegraphics[width=60mm,angle=270]{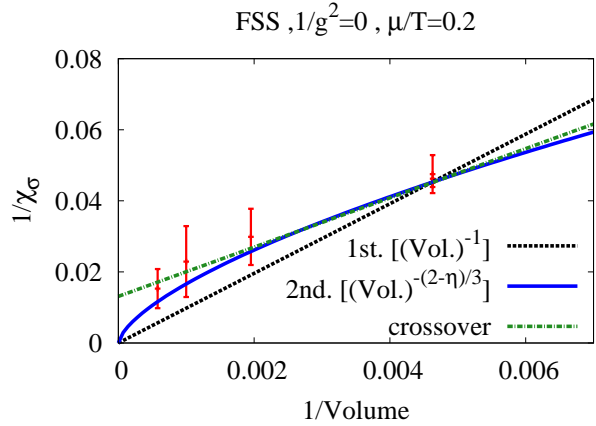}
  \end{center}
\caption{Finite size scaling of chiral susceptibility for $\mu /T=0.2$.
Dotted, solid, and dashed lines show expected behavior of the first, second, and crossover transitions.
We show chiral susceptibilities on the $6^3 \times 4, 6^3 \times 6, 8^3 \times 8, 10^3 \times 4$, and $12^3 \times 4$ lattices.
%\com{We omit results on a $4^3 \times 4$ lattice since $\Delta f$ in \figref{Fig:df} suggests that all the results except for $4^3 \times 4$ lattice results are scaled, and the $4^3 \times 4$ lattice size is not large enough to investigate finite size scaling.}
}
\label{Fig:FSS}
\end{figure}
%%%%%%%%%%%%%%%%%%%%%%%%%%%%%%%%%%%%%%%%%%%%%%%%%%%%%%%%%%%%%%%%%%%%%%%%%%%%%%%%
In order to find the finite size scaling of the chiral susceptibility~\cite{FSS},
we plot $1/\chi_\sigma$ at the peak as 
%a function 
\com{functions} of inverse spatial lattice volume
in \figref{Fig:FSS}.
The chiral susceptibility is proportional to spatial volume $V=L^3$
in the first order phase transition region 
and to 
%volume of the power of $(2-\eta)/3$ 
$V^{(2-\eta)/3}$
in the second order phase transition region
for $d=3$ O(2) spin systems,
where the O(2) critical exponent is $\eta = 0.0380(4)$ \cite{criticalexp}.
By comparison, $\chi_\sigma$ does not diverge
when the transition is crossover.
It seems to suggest
that the chiral phase transition at low $\mu$ is not the first order,
and we cannot exclude the possibility of the crossover transition
with the present precision
in comparison with the above three scaling functions
shown in \figref{Fig:FSS} in AFMC.
The current analysis implies that the phase transition
is the second order or crossover phase transition.
In order to conclude the order of the phase transition firmly,
we need higher-precision and larger volume calculations.

\subsection{High momentum mode contributions} 
\label{subsec:Highmode}
We quantitatively examine the influence of high momentum auxiliary field modes
on the average phase factor and the order parameters. 
For this purpose,
we compare the results by cutting off high momentum auxiliary field modes
having $\sum_j \sin^2 k_j > \Lambda$, where $\Lambda$ is a cutoff parameter.
The parameter $\Lambda$ is varied in the range $0 \leq \Lambda \leq d=3$
to examine their cutoff effects;
we include all Monte-Carlo configurations for $\Lambda =3$,
while we only take account of the lowest momentum modes for $\Lambda=0$. 

%%%%%%%%%%%%%%%%%%%%%%%%%%%%%%%%%%%%%%%%%%%%%%%%%%%%%%%%%%%%%%%%%%%%%%%%%%%%%%%%
%  figure
%%%%%%%%%%%%%%%%%%%%%%%%%%%%%%%%%%%%%%%%%%%%%%%%%%%%%%%%%%%%%%%%%%%%%%%%%%%%%%%%
\begin{figure*}[tbh]
    \begin{center}
      \PSfig{7.5cm}{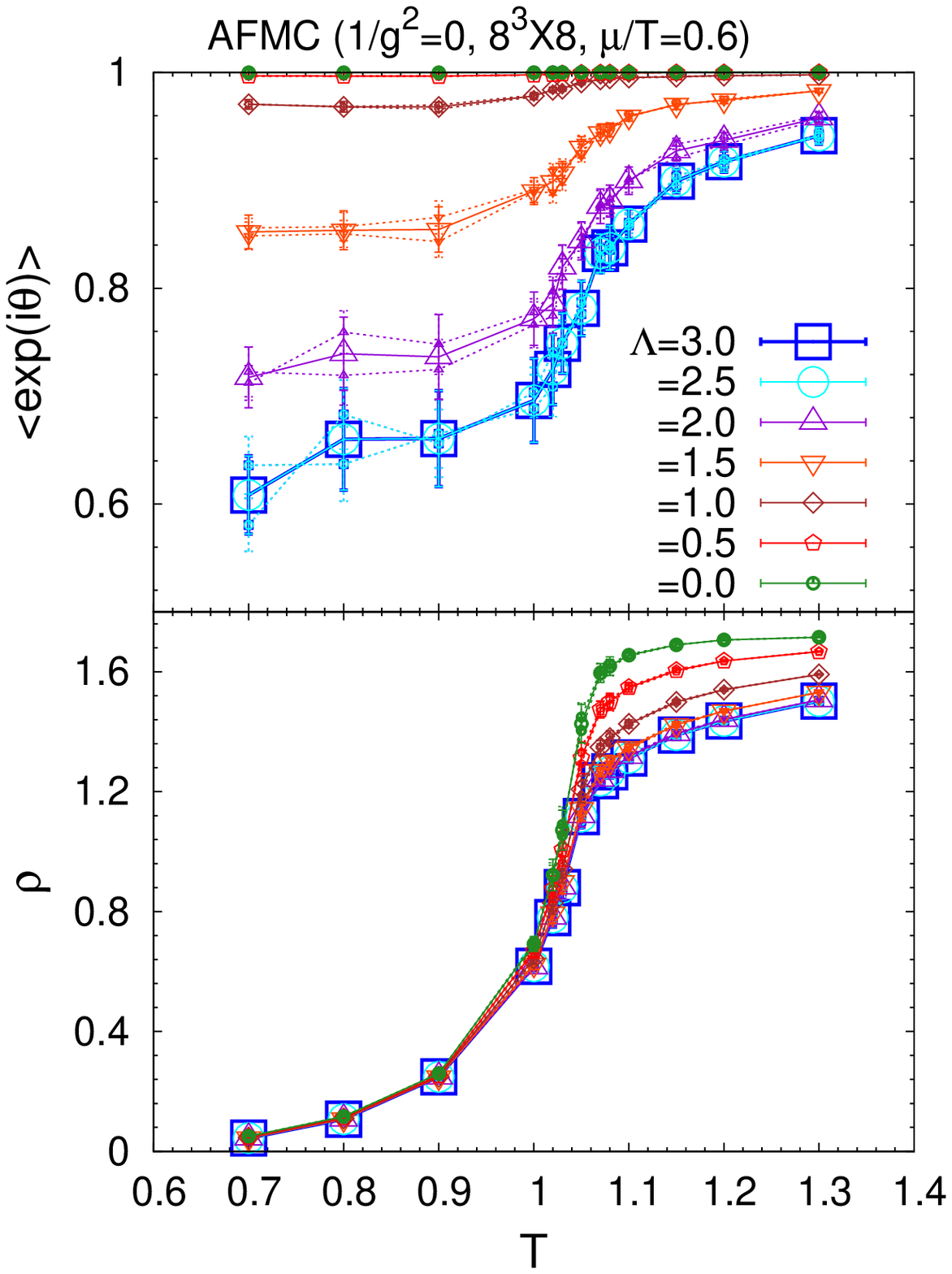}%
      \PSfig{7.5cm}{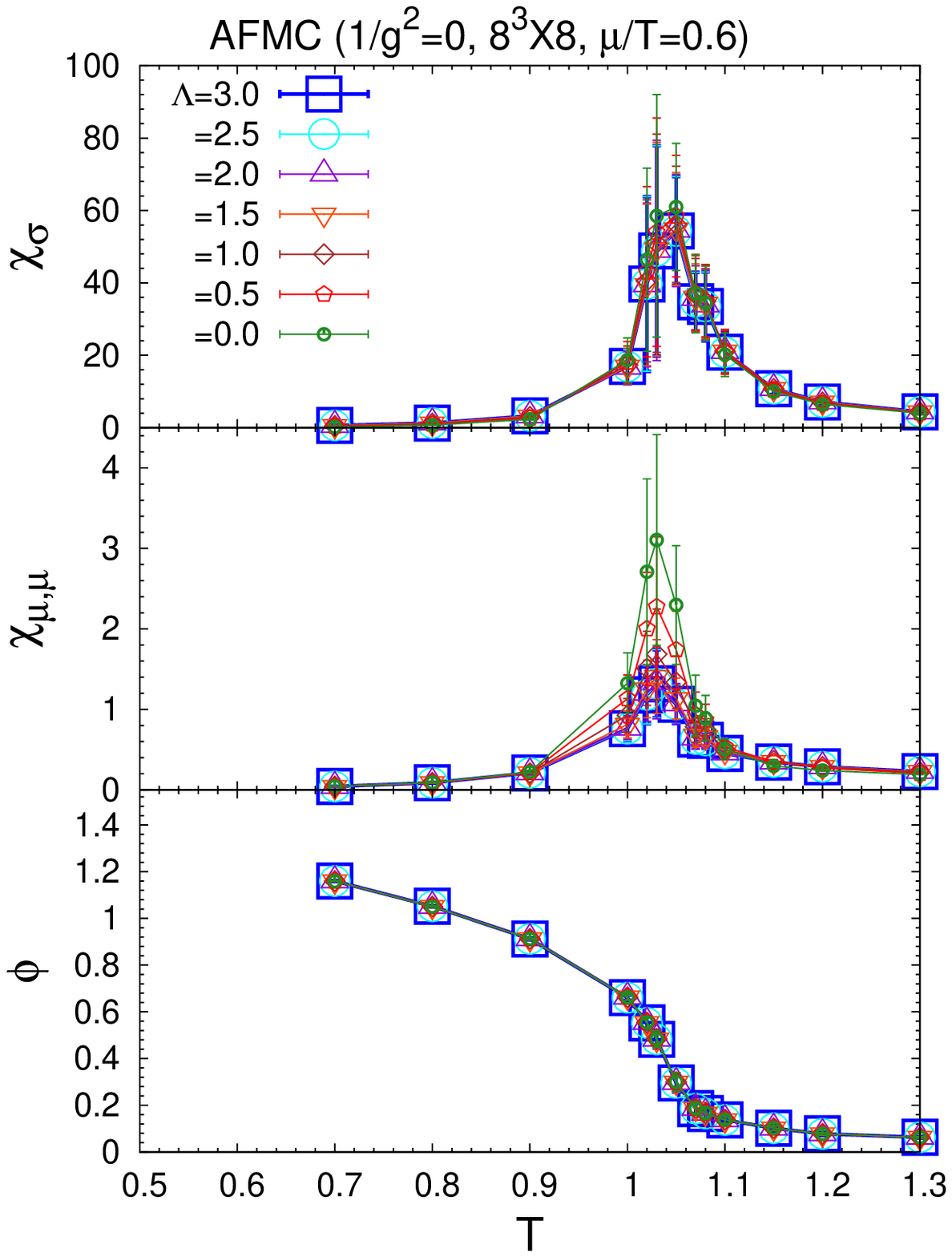}
    \end{center}
\caption{Cutoff parameter $\Lambda$ dependence as a function of temperature $T$
on a $8^3 \times 8$ lattice on a fixed $\mu /T=0.6$ line in the chiral limit.
We show the average phase factor (left top), quark number density (left bottom), chiral susceptibility (right top), quark number susceptibility (right middle),
and  \com{chiral condensate} (right bottom). 
Squares, big open circles, triangles, upside-down triangles, diamonds, pentagons and small filled circles show results for $\Lambda=3,2.5,2,1.5,1,0.5$ and 0, respectively. }
\label{Fig:cut}
\end{figure*}
%%%%%%%%%%%%%%%%%%%%%%%%%%%%%%%%%%%%%%%%%%%%%%%%%%%%%%%%%%%%%%%%%%%%%%%%%%%%%%%%
%  end figure
%%%%%%%%%%%%%%%%%%%%%%%%%%%%%%%%%%%%%%%%%%%%%%%%%%%%%%%%%%%%%%%%%%%%%%%%%%%%%%%%

The average phase factor might become large
in the cases where high momentum mode contributions
are negligible as discussed in \secref{subsec:Formalism}, 
so we anticipate that the weight cancellation becomes weaker 
for smaller $\Lambda$.
In the left top panel of Fig.~\ref{Fig:cut}, 
we show the $\Lambda$ dependence of the average phase factor 
on a $8^{3} \times 8$ lattice for $\mu /T=0.6$.
The average phase factor has a large value when $\Lambda \rightarrow 0$,
where we improve the weight cancellation.  
These results are consistent with our expectation 
for the weight cancellation with high momentum modes.
We could here conclude that high momentum modes are 
closely related to severe weight cancellation. 

In the right bottom panel of Fig.~\ref{Fig:cut},
we show the chiral condensate $\phi$
on a $8^3 \times 8$ lattice for $\mu /T=0.6$.
We here utilize $\phi = \expv{\sum_{\tau} \sigma_{\bm{k}=\bm{0},\tau} / \Nt }$.
This expression is equivalent to \equref{Eq:defsigma} for $\Lambda =3$.
The chiral condensate does not depend on the parameter $\Lambda$ 
since the lowest modes of the integration variables ($\sigma_\vkt,\  \pi_\vkt $) in AFMC consist of the scalar and pseudoscalar modes.
In Fig.~\ref{Fig:cut}, 
we also plot the cutoff dependence of other quantities: 
quark number density ($\rho_q$),
chiral susceptibility ($\chi_{\sigma}$)
and quark number susceptibility ($\chi_{\mu, \mu}$). 
We find that these quantities do not strongly
depend on the cutoff as long as $\Lambda \geq 2$. 
By contrast, the quantities are affected by the cutoff parameter for $\Lambda <2$.
We have already found that the average phase factor becomes large
if we set $\Lambda \leq 2.5$. 
Thus, this analysis implies a probable presence of an optimal
cutoff $\Lambda_{\mrm{o}}$, 
with which the order parameter values 
are almost the same as those of the all momentum modes 
and the reliability of numerical simulation 
is improved.
We conclude that there is a possible way to study the QCD phase diagram
for larger lattice by cutting off or approximating the high momentum modes
without changing the behavior of the order parameters.

\section{Summary}
\label{sec:SD}
We have investigated the QCD phase diagram 
and the sign problem
in the auxiliary field Monte-Carlo (AFMC)
method with chiral angle fixing (CAF) technique.
In order to obtain the auxiliary field effective action, 
we first integrate out spatial link variables and obtain
an effective action as a function of quark fields and temporal link variables  
in the leading order of the $1/g^2$ and $1/d$ expansion
with one species of unrooted staggered fermion.
%We next obtain 
By using the extended Hubbard-Stratonovich (EHS) transformation, 
we convert the four-Fermi interactions
into the bilinear form of quarks.
The auxiliary field effective action is obtained 
after analytic integration over the quark and temporal link variables.
%Finally, we integrate out the auxiliary fields
We have performed the auxiliary field integral using the Monte-Carlo technique.
%------------------------------------------------------------------------------*

We have obtained auxiliary field configurations in AFMC
and the order parameters:
the chiral condensate and quark number density.
Both of \com{these} order parameters show phase transition behavior.
In the low chemical potential region, 
the chiral condensate decreases smoothly with increasing temperature, 
while the quark number density increases gently.
This behavior suggests that 
\com{the phase transition is the second order} or crossover, 
which is consistent with the analysis of the distribution of the chiral condensate.
We call the low chemical potential region the would-be second order region.
In order to deduce the phase boundary, 
we here define (pseudo-)critical temperature as a peak position of the chiral susceptibility.
One finds that the critical temperature is suppressed compared with the mean-field results on an isotropic lattice~\cite{MF-SCL,NLOPD,NNLO} 
and \com{is} almost independent of lattice size as shown in the monomer-dimer-polymer simulations (MDP) 
at the would-be second order phase transition~\cite{MDP,PhDFromm}.
We also give some results of finite size scaling to guess the phase transition order.
While one could expect the second order phase transition from the mean-field and O(2) symmetry arguments in the low chemical potential region,
it is not yet conclusive to decide whether the transition is the second order or crossover at the present precision.

At high chemical potential,
the order parameters show sudden jump and hysteresis,
 and depend on initial conditions: the Wigner and Nambu-Goldstone initial conditions.
The distribution of the chiral condensate has a double peak around the phase transition region.
These results imply that the order of the phase transition is the first order 
owing to the existence of the two local minima 
with a relatively high barrier compared to the Metropolis jumping width.
We call this phase transition the would-be first order phase transition in the present paper.
We here regard transition temperature as a crossing point of the expectation value of the effective action with two initial conditions.
According to our analysis, the Nambu-Goldstone phase is enlarged toward the high chemical potential region compared with the mean-field results.
The phase boundary depends very weakly on spatial lattice size 
and more strongly on temporal lattice size.
This behavior is also found in MDP~\cite{MDP}.

We find that we have a sign problem in AFMC.
The origin of the weight cancellation 
is the bosonization of the negative modes in
the extended Hubbard-Stratonovich (EHS) transformation;
an imaginary number must be introduced in the fermion matrix.
The fermion determinant becomes complex, and 
the weight cancellation arises 
when we numerically integrate auxiliary fields.
In our framework, we have a phase cancellation mechanism for low momentum auxiliary fields; 
a phase on one site is canceled out by the nearest neighbor site phase.
We quantitatively show that the high momentum modes contribute
to the weight cancellation by cutting off these modes.
We also confirm the cutoff dependence on order parameters
and susceptibilities.
We find that there is a cutoff parameter region 
where the behavior of the quantities are not altered 
from the all mode results and the
weight cancellation is weakened.
Therefore, there is a possibility to investigate phase transition phenomena
using cutoff or approximation scheme for high momentum modes.

While we have a sign problem in AFMC,
\com{the} weight cancellation is not serious 
on small lattices adopted here ($\sim 8^3 \times 8$ size) 
because of the phase cancellation mechanism
for the low momentum modes.
The phase boundary in AFMC is found to be consistent
with that in MDP~\cite{MDP}.

In this paper, we utilize CAF 
in order to obtain the order parameters and susceptibilities in the chiral
limit on a fixed finite size lattice.
The chiral condensate in finite volume should vanish in a rigorous sense
due to the chiral symmetry
between the scalar and pseudoscalar modes.
In order to simulate the non-vanishing chiral condensate to be
obtained in the rigorous procedure of the thermodynamic limit 
followed by the chiral limit,
the chiral transformation of auxiliary fields are carried out
in each configuration
so as to fix the chiral angle \sout{to be} in the real positive direction (positive scalar mode direction).
We could evaluate the adequate chiral condensate and chiral susceptibility by using CAF.

The AFMC method could be straightforwardly applied 
to include finite coupling effects
since bosonization technique is applied in the mean-field analysis~\cite{NLOPD,NNLO}.
Both fluctuations and finite coupling effects are important to elucidate features of the phase transition phenomena, 
so the AFMC would be a possible way to include these two effects at a time.
The sign problem might be severer than that in the strong coupling limit
when we include finite coupling effects. 
One of the promising methods to avoid lower numerical reliability
is to invoke shifted contour formulation~\cite{avoidsign},
and another promising direction is to integrate 
over the Lefschetz thimbles~\cite{thimble}.
We hope that we may apply the formulation
with finite coupling effects or on a larger lattice.
We also obtain appropriate order parameters 
in a relatively hassle-free CAF method
compared to a rigorous way.
In AFMC, chiral symmetry is respected manifestly.
We expect that auxiliary fields for finite coupling terms also keep chiral symmetry manifestly, and CAF can be also applied to finite coupling cases.
We might use this CAF method with higher-order corrections in the strong coupling expansion to investigate the phase diagram.
It is also important to develop a method to include spatial baryon hopping terms and other higher order terms in the $1/d$ expansion in AFMC.

\section*{Acknowledgment}
The authors would like to thank Wolfgang Unger, Philippe de Forcrand, 
Naoki Yamamoto, Kim Splittorff, Jan M. Pawlowski, Mannque Rho, Atsushi Nakamura,
Hideaki Iida, Sinya Aoki, Frithjof Karsch, Swagato Mukherjee 
and participants of the YIPQS-HPCI workshop
on "New-type of Fermions on the Lattice"
and the YIPQS workshop on "New Frontiers in QCD 2013" (YITP-T-13-05)
for useful discussions.
TI is supported by the Grants-in-Aid for JSPS Fellows (No.25-2059). 
TZN is supported by the Grant-in-Aid for JSPS Fellows (No.22-3314).
This work is supported in part by the Grants-in-Aid for Scientific Research
from JSPS
(Nos.
          %09J01226, %(T. Misumi)
          %10J03314, %(T.Z. Nakano)
          %11J00593, %(T. Kimura)
          (A) 23340067, %(T. Kunihiro (incl. A.Ohnishi))
          (B) 24340054, %(A. Nakamura (incl. A.Ohnishi))
          (C) 24540271%(A. Ohnishi, K. Morita, T. Kunihiro)
),
by the Grants-in-Aid for Scientific Research on Innovative Areas from MEXT
(No. 2404: 24105001, 24105008), % AO (NS: X01, D01)
by the Yukawa International Program for Quark-hadron Sciences,
and by the Grant-in-Aid for the global COE program ``The Next Generation
of Physics, Spun from Universality and Emergence" from MEXT.

\appendix
\section{Temporal-spatial lattice spacing ratio on an anisotropic lattice}
\label{App:fgamma}
\com{
\com{In this paper,} 
we have adopted the physical temporal-spatial lattice spacing ratio
$f(\gamma)=a_s^{\mrm{phys}}/a_\tau^{\mrm{phys}}=\gamma^2$
given in Refs.~\cite{Bilic,BilicDeme,PhDFromm}.
%~\cite{Bilic,BilicDeme,PhDFromm}.
We briefly summarize the arguments given in these references.
% Refs.~\cite{Bilic,BilicDeme,PhDFromm}.

In general, we can set temporal lattice spacing 
as $a_\tau^{\mrm{phys}} (\gamma , g)=a_s^{\mrm{phys}}/f(\gamma, g)$, 
where $f(\gamma,\infty)=f(\gamma)$ in SCL.
In order to give the expression for the function $f$,
we can \com{impose}\sout{request} \com{\sout{set} two conditions}
that critical temperature 
$T_c=1/N_\tau a_\tau^\mrm{phys}=f(\gamma)/N_\tau a_{\com{s}}^\mrm{phys}$
should not depend on $N_\tau$,
and that the hypercube symmetry is restored on an isotropic lattice,
 $a_\tau^{\mrm{phys}} (1,g)=a_s^{\mrm{phys}}$.
Since the critical coupling $\gamma_c$ at $\mu=0$ is given as
$\gamma^2_c/N_\tau = d(N_c+1)(N_c+2)/6(N_c+3)$
for $\mathrm{SU}(N_c)$
with one species of unrooted staggered fermion 
in the mean field treatment in SCL,
the above requirements
suggest that the function $f$ should be given
as $f(\gamma)=\gamma^2$.

This prescription seems to be
valid also for high chemical potential region 
as long as $\gamma$ is bigger than unity.
This is because critical chemical potential $\mu_c$ 
at zero temperature becomes constant as long as $\gamma \gg 1$ 
when we apply $f(\gamma) = \gamma^2$,
which is consistent with the zero chemical potential result.
%We follow this argument and adopt $f(\gamma)=\gamma^2$.
}

\section{Chiral Angle Fixing} 
\label{App:CAF}
\begin{figure}[tbh]
    \begin{center}
       \includegraphics[width=100mm]{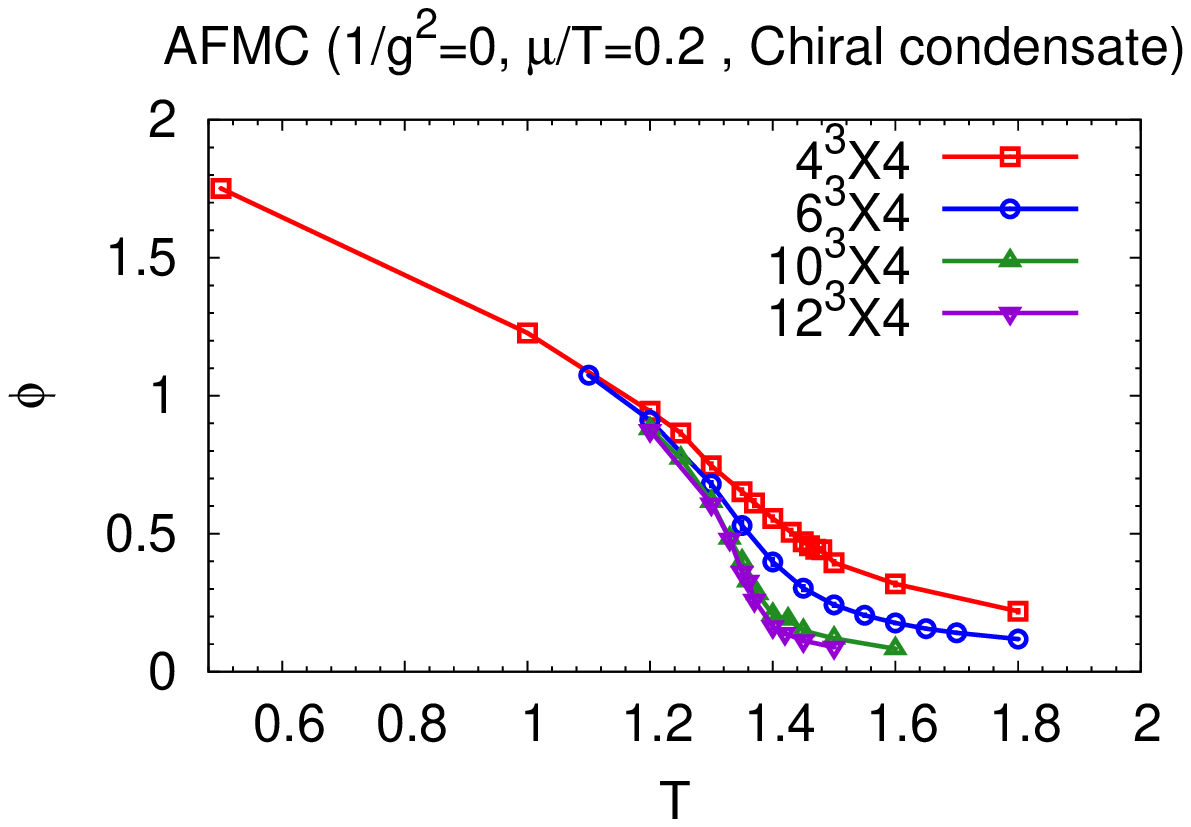}
    \end{center}
\caption{
Spatial lattice size dependence of the chiral condensate on the %\sout{fixed}
$\mu /T =0.2$ line.
In the Wigner phase, 
the chiral condensate %\sout{tends to be zero as}
decreases with
increasing spatial lattice size.}
\label{Fig:csigma}
\end{figure}

We here discuss the chiral angle fixing (CAF) from 
another point of view.
As mentioned in \secref{Sec:CAF}, 
the chiral condensate ideally disappear\com{s} due to the chiral symmetry.
We could confirm an aspect of CAF method as follows.
According to a relation, 
$S(\phi_{\bk, \omega},\alpha_{\bk,\omega},\phi,\alpha)
=S(\phi_{\bk, \omega},\alpha_{\bk,\omega}'=\alpha_{\bk,\omega}-\alpha,\phi,0)$,
where $\alpha$ is the chiral angle,
the chiral condensate is given as
\begin{align}
\expv{\sigma_{0}} =& 
\frac{1}{Z}\int \mathcal{D} \left[ \sigma_{0}, \pi_{0}\right]\ 
\prod_{(\bk ,\omega)\not=(\bm{0},0)}
%\prod_{\sigma_{\bk,\omega},\pi_{\bk,\omega}}
\mathcal{D}\left[ \sigma_{\bk,\omega}, \pi_{\bk,\omega}\right]\ 
\sigma_{0} \ 
%\nonumber\\
%\times
e^{-S(\sigma_{\bk, \omega},\pi_{\bk,\omega},
\phi=\sqrt{\sigma_{0}^{2}+\pi_{0}^{2}},\alpha)}\ \nonumber \\
=& 
\frac{1}{Z}\int \mathcal{D} \left[ \phi, \alpha\right]\ \phi \cos\alpha 
%\nonumber\\
%\times
\int 
\prod_{(\bk ,\omega)\not=(\bm{0},0)}
\mathcal{D} \left[\phi_{\bk,\omega}, \alpha_{\bk,\omega}'\right] 
\ e^{-S(\phi_{\bk, \omega},\alpha_{\bk,\omega}',\phi,0)} 
\nonumber \\
=& 0
\ .
\label{Eq:bCAF}
\end{align}
$\phi_{\bk, \omega}$ and $\alpha_{\bk,\omega}$ are chiral radius 
and chiral angle of each chiral partner.
We find that the chiral condensate ideally vanishes according to \equref{Eq:bCAF}. 
In CAF,
we rotate the negative chiral angle ($-\alpha$) with respect to all fields and set $\pi_0 =0$. We obtain the finite chiral condensate in the Nambu-Goldstone (NG) phase as
\begin{align}
\expv{\sigma_0 } 
=& \frac{1}{Z} \int \mathcal{D} \sigma_0 \ \sigma_{0} 
%\prod_{\phi_{\bk ,\omega},\alpha_{\bk,\omega}}
%\nonumber\\
%\times&
\int
\prod_{(\bk ,\omega)\not=(\bm{0},0)}
\sout{\int} 
\mathcal{D} \left[ \phi_{\bk,\omega},\alpha_{\bk,\omega} \right] \ 
\ e^{-S(\phi_{\bk, \omega},\alpha_{\bk,\omega},\phi=\sigma_{0},0)}
\nonumber\\
\neq& 0 \ .
\label{Eq:aCAF}
\end{align}
The resultant chiral condensate in CAF 
should simulate the spontaneously broken chiral condensate
in the thermodynamic limit.
\com{
In the Wigner phase, $\left< \sigma_0 \right>$ after CAF takes a finite value on the finite size lattice, but it decreases with
increasing lattice size
as shown in \figref{Fig:csigma}.
In the limit of infinite volume, we could expect that the chiral condensate 
approaches zero in the Wigner phase, as expected
in the continuum theories.
}
%%%%%%%%%%%%%%%%%%%%%%%%%%%%%%%
%  figure
%%%%%%%%%%%%%%%%%%%%%%%%%%%%%%%%%%%%%%%%%%%%%%%%%%%%%%%%%%%%%%%%%%%%%%%%%%%%%%%%
%%%%%%%%%%%%%%%%%%%%%%%%%%%%%%%%%%%%%%%%%%%%%%%%%%%%%%%%%%%%%%%%%%%%%%%%%%%%%%%%
%  end figure
%%%%%%%%%%%%%%%%%%%%%%%%%%%%%%%%%%%%%%%%%%%%%%%%%%%%%%%%%%%%%%%%%%%%%%%%%%%%%%%%

We have some advantages in CAF.
One is that the chiral condensate is finite in the NG phase
and the chiral susceptibility may have a peak.
In the cases where the chiral condensate vanishes ($\expv{\sigma_{0}} =0$),
%because of the chiral symmetry,
the chiral susceptibility, 
%which is proportional to $\partial^2 \ln Z /\partial m_0^2 =-\partial \expv{\chibar \chi}/ \partial m_0 = \partial \expv{\sigma_0}/ \partial m_0 $, 
%is expressed as %$\chi_{m_0,m_0}=$
%$\partial^2 \ln Z /\partial m_0^2 =$
%$\expv{\sigma_0^2}$.
$\chi_{\sigma} = \expv{\sigma_0^2}- \expv{\sigma_0}^2$,
becomes $\chi_{\sigma} \rightarrow \expv{\sigma_0^2}$, 
then
we could expect that the chiral susceptibility increases with lower temperature.
After we utilize CAF, 
we obtain the chiral susceptibility with a peak 
%$\partial^2 \ln Z /\partial m_0^2 =\expv{\sigma_0^2}- \expv{\sigma_0}^2$ 
because of the non-vanishing chiral condensate at low $T$ as shown in \figref{Fig:chicom}. 
Another merit of CAF is that when we calculate the chiral condensate and the chiral susceptibility, 
we could take into account the information on the pseudoscalar mode which is mixed with scalar mode in the chiral limit.
%\begin{widetext}

\section{Recursion formula, Order parameters \& Susceptibilities} \label{sec:rec}
\subsection{Recursion formula}
First, we shortly introduce a matrix to confirm notation \cite{Faldt,Bilic,BilicDeme}.
The function, $B_{\Nt}$, is defined as 
%%%%%%%%%%%%%%%%%%%
\begin{eqnarray}
B_{\Nt}(I_{1},\cdots,I_{\Nt}) =
\left| 
\begin{array}{ccccc}
I_1  & e^{\mu/\gamma^2}  & 0 & \cdots & 0 \\
-e^{-\mu/\gamma^2}  & I_2 & e^{\mu/\gamma^2}  &  & \vdots \\
0 & -e^{-\mu/\gamma^2}  & I_3 \cdot & \ddots & \vdots \\
\vdots & 0 & \ddots & \ddots & 0 \\
\vdots & \ddots & \ddots & I_{\Nt -1} & e^{\mu/\gamma^2}  \\
0 & \cdots & 0 & -e^{-\mu/\gamma^2}  & I_{N_\tau} \\
\end{array} 
\right| 
\ , \ \label{Eq:Bn}
\end{eqnarray}
%%%%%%%%%%%%%%%%%%%
where $I_{x}=2m_{x}/\gamma$.
In the mean-field approximation, $I_i = \mathrm{const.}$ for any $i$.
By comparison, $I_i$ depends on the spacetime 
when fluctuations are taken into account,
therefore $I_i \neq I_j$ for $i \neq j$. 
Using \equref{Eq:Bn}, 
$B_{\Nt}$ can be written in recursion formulae~\cite{Faldt,Bilic,BilicDeme},

\begin{align}
\label{Eq:recB_N}
B_{\Nt} \left(I_1,\cdots,I_{\Nt} \right) 
&= I_{\Nt}B_{\Nt-1}\left(I_1,\cdots,I_{\Nt-1} \right)  
+ B_{\Nt-2} \left(I_1,\cdots,I_{\Nt-2} \right) 
\ ,
\end{align}
where $B_1=I_1$, $B_2=I_1I_2+1$.
$X_{N_\tau}$ in \equref{Eq:SeffAF} is obtained by using $B_{\Nt}$ as
\begin{align}
\label{Eq:XN}
X_{\Nt} \left(I_1,\cdots,I_{\Nt} \right)
&= B_{\Nt} \left(I_1,\cdots,I_{\Nt} \right)
+ B_{\Nt -2}\left( I_2,\cdots,I_{\Nt -1}\right) 
\ . 
\end{align}

In this paper, we numerically calculate $X_{\Nt}$ and $B_{\Nt}$ by 
\com{using} these recursion formulae.

\subsection{Order parameters and Susceptibilities}
We here summarize the expression for order parameters and susceptibilities \cite{Faldt}.
The chiral condensate \com{$\langle$} $\sigma_0$ \com{$\rangle$} and the chiral susceptibility $\chi_{\sigma}$ are described as
%*************************
\begin{align}
\expv{\sigma_{0}} =& -\frac{1}{L^3 N_{\tau}}\frac{\partial \ln Z}{\partial m_{0}} 
%\nonumber \\
=- \frac{1}{L^3 N_{\tau}}\frac{1}{Z} \int \mathcal{D} \left[ \sigma, \pi \right] \left(-\frac{\partial S_{\mrm{eff}}^{\mrm{AF}}}{\partial m_{0}} \right) e^{-S_{\mrm{eff}}^{\mrm{AF}}} 
%\nonumber \\
=\frac{1}{L^3 N_{\tau}} \expv{\frac{\partial S_{\mrm{eff}}^{\mrm{AF}}}{\partial m_{0}}} 
\ , \label{Eq:appsigma}
 \\
\chi_{\sigma}=& \frac{1}{L^3 N_{\tau}}\frac{\partial^2 \ln Z}{\partial m_{0}^2}
% \nonumber \\
= \frac{1}{L^3 N_{\tau}}\left[ \expv{\left( \frac{\partial S_{\mrm{eff}}^{\mrm{AF}}}{\partial m_0}\right)^2 } -\expv{\frac{\partial S_{\mrm{eff}}^{\mrm{AF}}}{\partial m_0}}^2 - \expv{\frac{\partial^2 S_{\mrm{eff}}^{\mrm{AF}}}{\partial m_0^2}}
\right] \nonumber \\
=& \frac{1}{L^3 N_{\tau}}\left[ \expv{\left( \frac{\partial S_{\mrm{eff}}^{\mrm{AF}}}{\partial m_0}  -\expv{\frac{\partial S_{\mrm{eff}}^{\mrm{AF}} }{\partial m_0}}\right)^2 } - \expv{\frac{\partial^2 S_{\mrm{eff}}^{\mrm{AF}}}{\partial m_0^2}} \right]
\ , \label{Eq:appchi} 
\end{align}
%**************************
where $Z=\int \mathcal{D} \left[ \sigma, \pi \right] \exp \left( -S_{\mrm{eff}}^{\mrm{AF}}\right)$.
The derivatives of the effective action in 
\equref{Eq:appsigma} \com{and}~(\ref{Eq:appchi}) are given as
%**************************
\begin{align}
\frac{\partial S_{\mrm{eff}}^{\mrm{AF}} }{\partial m_0} 
=& -\sum_{\bm{x}} \frac{\partial \ln \mathcal{\mathcal{K}}}{\partial m_0} 
= -\sum_{\bm{x}} \frac{1}{\mathcal{K}}\frac{\partial I_x}{\partial m_0} \frac{\partial X_{\Nt}}{\partial I_x}\frac{\partial \mathcal{K}}{\partial X_{\Nt}} \nn \\
=&-\sum_{\bm{x}}\frac{1}{\mathcal{K}}\frac{2}{\gamma}\sum_{t}B_{\Nt-1}\left(I_{t+1},\cdots,I_{\Nt},I_1,\cdots,I_{t-1} \right) \left( 3X_{\Nt}^2-2 \right)
\ , \\
\frac{\partial^2 S_{\mrm{eff}}^{\mrm{AF}} }{\partial m_0^2} 
=&  \sum_{\bm{x}} \left[ \frac{1}{\mathcal{K}^2} \left(\frac{2}{\gamma}\sum_{t}B_{\Nt-1}\left(I_{t+1},\cdots,I_{\Nt},I_1,\cdots,I_{t-1} \right)\left( 3X_{\Nt}^2-2 \right) \right)^2 \right.
 \nn \\
& -\frac{1}{\mathcal{K}} \left(\frac{2}{\gamma} \right)^2 \sum_{t,t'}  
  \left\{ \sum_{t > t'}  B_{N-t+t'-1}(I_{t+1},\cdots,I_{N},I_{1},\cdots,I_{t'-1})B_{t-t'-1}(I_{t'+1},\cdots,I_{t-1}) \right. 
\nonumber \\
& \left. +\sum_{t<t'} B_{t'-t-1} (I_{t+1},\cdots,I_{t'-1})B_{N-t'+t-1}(I_{t'+1},\cdots,I_{N},I_{1},\cdots,I_{t-1})\right\}  \left( 3X_{\Nt}^2-2 \right) 
 \nn \\
& \left. - \frac{1}{\mathcal{K}}\frac{24}{\gamma^2} X_{\Nt} \left( \sum_{t} B_{\Nt-1}\left(I_{t+1},\cdots,I_{\Nt},I_1,\cdots,I_{t-1} \right)  \right)^2  \right]
\ ,
\end{align}
%**************************
where $\mathcal{K}=X_{\Nt}^3 -2X_{\Nt}+2\cosh (3\Nt \mu /\gamma^2)$.
Similarly, we also obtain quark number density $\rho_q$, quark number susceptibility $\chi_{\mu,\mu}$ and mixed susceptibility $\chi_{m_0 ,\mu}$ 
%**************************
\begin{align}
\rho_q =& -\frac{T}{L^3}\frac{\partial \ln Z}{\partial \mu} 
%\nonumber \\
%=- \frac{1}{Z} \int \mathcal{D} \left[ \sigma, \pi \right] \left(-\frac{\partial S_{\mrm{eff}}^{\mrm{AF}}}{\partial \mu} \right) e^{-S_{\mrm{eff}}^{\mrm{AF}}} 
%\nonumber \\
=\frac{T}{L^3} \expv{\frac{\partial S_{\mrm{eff}}^{\mrm{AF}}}{\partial \mu}}
\ , \\
\chi_{\mu,\mu}=& \frac{1}{L^3 N_{\tau}}\frac{\partial^2 \ln Z}{\partial \mu^2 }
% \nonumber \\
%= \expv{\left( \frac{\partial S_{\mrm{eff}}^{\mrm{AF}}}{\partial \mu}\right)^2 } -\expv{\frac{\partial S_{\mrm{eff}}^{\mrm{AF}}}{\partial \mu}}^2 - \expv{\frac{\partial^2 S_{\mrm{eff}}^{\mrm{AF}}}{\partial \mu^2}}
%\nonumber \\
= \frac{1}{L^3 N_{\tau}}\left[ \expv{\left( \frac{\partial S_{\mrm{eff}}^{\mrm{AF}}}{\partial \mu}  -\expv{\frac{\partial S_{\mrm{eff}}^{\mrm{AF}}}{\partial \mu}}\right)^2 } - \expv{\frac{\partial^2 S_{\mrm{eff}}^{\mrm{AF}}}{\partial \mu^2}} \right]
\ ,\\
%\end{align}
%\begin{align}
\chi_{m_0,\mu}=& \frac{1}{L^3 N_{\tau}} \frac{\partial^2 \ln Z}{\partial \mu \partial m_0}
% \nonumber \\
= \frac{1}{L^3 N_{\tau}} \left[ \expv{ \frac{\partial S_{\mrm{eff}}^{\mrm{AF}}}{\partial m_0} \cdot \frac{\partial S_{\mrm{eff}}^{\mrm{AF}}}{\partial \mu} } -\expv{\frac{\partial S_{\mrm{eff}}^{\mrm{AF}}}{\partial m_0}}\expv{\frac{\partial S_{\mrm{eff}}^{\mrm{AF}}}{\partial \mu}} - \expv{\frac{\partial^2 S_{\mrm{eff}}^{\mrm{AF}}}{\partial m_0 \partial \mu}} \right]
\ ,
\end{align}
%*************************
where
%***********************************************************************
\begin{align}
\frac{\partial S_{\mrm{eff}}^{\mrm{AF}} }{\partial \mu} 
 =& -\sum_{\bm{x}} \frac{\partial \ln \mathcal{\mathcal{K}}}{\partial \mu} 
= -\sum_{\bm{x}} \frac{1}{\mathcal{K}} \frac{2\cdot 3\Nt}{\gamma^2} \sinh \left(\frac{3\Nt \mu}{\gamma^2}\right) 
\ , \\
\frac{\partial^2 S_{\mrm{eff}}^{\mrm{AF}} }{\partial \mu^{2}} 
=& \sum_{\bm{x}} \left[ \frac{1}{\mathcal{K}^2} \left( \frac{2\cdot 3 \Nt}{\gamma^2} \sinh\left(\frac{3\Nt \mu}{\gamma^2}\right)    \right)^2 - \frac{2}{\mathcal{K}}\left( \frac{ 3\Nt }{\gamma^2} \right)^2 \cosh \left(\frac{3\Nt \mu}{\gamma^2}\right) \right]
\ , \\
\frac{\partial^2 S_{\mrm{eff}}^{\mrm{AF}} }{\partial m_0 \partial \mu} 
=& \sum_{\bm{x}} \frac{1}{\mathcal{K}^2} \frac{2\cdot 3 \Nt}{\gamma^2} \sinh \left( \frac{3\Nt \mu}{\gamma^2}\right) \left( \frac{2}{\gamma} \right)
\nn \\
& \times \sum_{t}B_{\Nt-1}\left(I_{t+1},\cdots,I_{\Nt},I_1,\cdots,I_{t-1} \right) \left( 3X_{\Nt}^2- 2 \right)
\ .
\end{align}
%***********************************************************************
%\end{widetext}


%\bibliographystyle{ptephy}
%\bibliography{sample}
%
% once the .bbl file has been generated then place the text in your article.

%\vfill\pagebreak

\end{document}